\documentclass[jgrga]{agutex}
 \pdfoutput=1

 \usepackage{lineno}

\usepackage{graphicx}
\usepackage{epstopdf}
\usepackage{appendix}
\usepackage{caption}
\setkeys{Gin}{draft=false}

\authorrunninghead{MISHRA ET AL.}

\titlerunninghead{Interacting CMEs of September 25 and 28, 2012}

\begin{document}

\title{Kinematics of interacting CMEs of September 25 and 28, 2012 }

%
%



 \authors{Wageesh Mishra \altaffilmark{1,2}, Nandita Srivastava \altaffilmark{1} and Talwinder Singh \altaffilmark{3}}

\altaffiltext{1}{Udaipur Solar Observatory, Physical Research Laboratory, Udaipur 313001, India}

\altaffiltext{2}{CAS Key Laboratory of Geospace Environment, Department of Geophysics and Planetary Sciences,
University of Science and Technology of China, Hefei, Anhui 230026, China}

\altaffiltext{3}{Indian Institute of Technology, Banaras Hindu University, Varanasi 221005, India}

%

\begin{abstract}
We have studied two Coronal Mass Ejections (CMEs) that occurred on September 25 and 28, 2012 and interacted near the Earth. By fitting the  Graduated Cylindrical Shell (GCS) model on the SECCHI/COR2 images and applying the Stereoscopic Self-Similar Expansion (SSSE) method on the SECCHI/HI images, the initial direction of both the CMEs is estimated to be west of the Sun-Earth line. Further, the three-dimensional (3D) heliospheric kinematics of these CMEs have been estimated using Self-Similar Expansion (SSE) reconstruction method. We show that use of SSE method with different values of angular extent of the CMEs, leads to significantly different kinematics estimates for the CMEs propagating away from the observer. Using the estimated kinematics and true masses of the CMEs, we have derived the coefficient of restitution for the collision which is found to be close to elastic. The in situ measurements at 1 AU show  two distinct structures of interplanetary CMEs, heating of the following CME, as well as ongoing interaction between the preceding and the following CME. We highlight the signatures of interaction in remote and in situ observations of these CMEs and the role of interaction in producing a major geomagnetic storm. 
\end{abstract}

\begin{article}

\section{Introduction}
\label{IntCMEsIntro}
The typical transit time of Coronal Mass Ejections (CMEs) from the Sun to the Earth is between 1 to 4 days and the number of CMEs launched from the Sun is about 5 per day around solar maximum \citep{StCyr2000,Yashiro2004,Webb2012}. Sometimes, CMEs occur in quick succession and under  favorable  conditions, they can interact or merge with each other as they propagate out in the heliosphere. The interaction of CMEs is more common near the solar maximum and therefore  the space weather prediction schemes are likely to be unsuccessful without considering the post-interaction CME characteristics.  Although the possibility of CME-CME interaction has been inferred earlier \citep{Intriligator1976,Burlaga1987}, the first observational remote evidence using  coronagraphic observations of Large Angle and Spectrometric COronagraph (LASCO; \citealp{Brueckner1995}) on-board \textit{SOlar and Heliospheric Observatory (SOHO)} and long wavelength radio observations,  was provided by \citet{Gopalswamy2001apj}. Also, the identification of a set of successive Earth-directed halo CMEs  and their appearance as complex ejecta in in situ observations near 1 AU has been reported  by \citet{Burlaga2001, Burlaga2002} and \citet{Wang2002a,Wang2003}.  It has been shown that CME-CME interactions are important as they can enhance the geoeffectiveness due to extended period and enhanced strength of southward magnetic field responsible generally for major geomagnetic storms \citep{Wang2003,Farrugia2004,Xue2005,Farrugia2006}.  Such interactions may help to understand the nature of collisions between magnetized plasmoids which are not expected to mix with each other, like ordinary gas bubbles. Also, if a shock from a following CME penetrates a preceding CME, it provides a unique opportunity to study the evolution of the shock strength, structure and its effect on plasma parameters as well as geoeffectiveness of preceding CME \citep{Wang2003a,Lugaz2005,Mostl2012,Liu2012,Lugaz2015}. These  information are crucial for predicting the arrival time of such  CMEs at the Earth, as the CME-CME interactions are expected to change the dynamics of interacting CMEs as they propagate out in the heliosphere.

Prior to the imaging observations of the inner heliosphere from \textit{Solar TErrestrial RElations Observatory} (\textit{STEREO}) \citep{Kaiser2008} to till date, several attempts to understand the CME-CME interaction have been  carried out based on magnetohydrodynamic (MHD) numerical simulations of the interaction of a shock wave with a magnetic cloud \citep{Vandas1997,Vandas2004,Xiong2006} and the interaction of two ejecta or magnetic cloud \citep{Gonzalez-Esparza2004,Lugaz2005,Wang2005,Xiong2007,Xiong2009,Lugaz2013,Shen2013,Shen2014}.  More recently \textit{STEREO} observations have enabled us to track a CME continuously from its lift-off in the corona up to the Earth and beyond. In particular the observations made  from the Heliospheric Imager (HI) on the SECCHI package \citep{Howard2008} and from their 3D stereoscopic reconstruction  from the twin vantage points, have provided direct evidence of CME-CME interaction in the inner heliosphere and beyond. A few cases of interacting CMEs have been studied recently using these observations. For example, the interacting CMEs of 2010 August 1 have been studied extensively by using primarily the \textit{STEREO}/HI (white light imaging), near Earth in situ and, \textit{STEREO}/Waves radio observations \citep{Harrison2012,Liu2012,Mostl2012,Temmer2012,Martinez-Oliveros2012,Webb2013}. Also, \citet{Lugaz2012} have reported a clear deflection of 2010 May 23-24 CMEs after their interaction.

Some of the key questions that have been recently addressed using \textit{STEREO} observations  is how the dynamics of CMEs change after interaction, what is the nature of collision, and how the plasma parameters of the CMEs are modified during their collision. The nature of collision of interacting CMEs have been investigated by various authors and is found that its regime can range from super-elastic to inelastic \citep{Lugaz2012,Shen2012,Shen2013,Mishra2014a,Temmer2014,Mishra2015}. These studies also show that collision has significant impacts on both the preceding and the following CMEs. The recent studies of CME-CME interaction have also addressed the question as to what is their role in enhancing the geoeffectiveness \citep{Mishra2014a,Mishra2015}.  These studies  show formation of an interaction region  as a result of collision of CMEs which is marked by a magnetic hole (probably, the signatures of magnetic reconnection).  These interaction regions seem to be responsible for major geomagnetic activity.

Recently, a study of the interacting CMEs of September 25 and 28, 2012, which occurred almost $\sim$ 2.5 days apart, has been reported by \citep{Liu2014}. In their study, they have focused on the merging of both the CMEs which began near the Earth and produced two step major geomagnetic storm. However, in the present  paper we  highlight on  the estimation of CME kinematics using Self-Similar Expansion (SSE) method developed by \citet{Davies2012}, identification of signatures of collision, nature of collision and role of interaction in producing a strong geomagnetic storm. Moreover, we have also examined the effect of considered angular width of the CME on its kinematics, estimated from SSE reconstruction method in which such a width is used as an input parameter.

\section{Observations and Analysis}
The CME of September 25, 2012 was launched from the Sun with a plane-of-sky speed of 335 km s$^{-1}$ and was found to accelerate slowly as observed in the LASCO/SOHO coronagraph images (CME catalog: $http://cdaw.gsfc.nasa.gov/CME\_list/$; see \citealp{Yashiro2004}). This CME was associated with a C1.1 class flare in the NOAA AR 11575 at the location W04N08 at 9:37 UT \citep{Liu2014}. On September 28, a full halo CME was launched at around 00 UT with a plane-of-sky speed of 950 km s$^{-1}$ which decelerated gradually. This CME was associated with a C3.7 flare in the NOAA AR 11577 at the location W30N09 on September 27, at 23:36 UT.

The kinematics of these selected CMEs have been estimated by (i) Graduated Cylindrical Shell (GCS) approximation of 3D reconstruction on COR2 images (ii) Construction of J-map (time-elongation plot) for the  HI images. Further, using a suitable single spacecraft reconstruction technique, namely the SSE method \citep{Davies2012}, on this time-elongation plot, the kinematics of the CMEs were estimated. The true masses of the two participating CMEs in interaction have also been estimated. Using the estimated kinematics before and after the collision and  their true masses, the coefficient of restitution has been calculated for the two CMEs, in order to understand their nature of collision. In-situ observations obtained from Wind \citep{Ogilvie1995} or \textit{Advanced Composition Explorer} (ACE; \citealp{Stone1998}) spacecraft for these interacting CMEs have also been examined, in order to understand the signatures and the role of interaction in enhancing the  geoeffectiveness of CMEs.

\subsection{3D reconstruction in COR2 field-of-view}
\label{3DCOR}

\begin{figure}
\captionsetup{width=0.85\textwidth}
\begin{center}
\includegraphics[angle=0,scale=0.35]{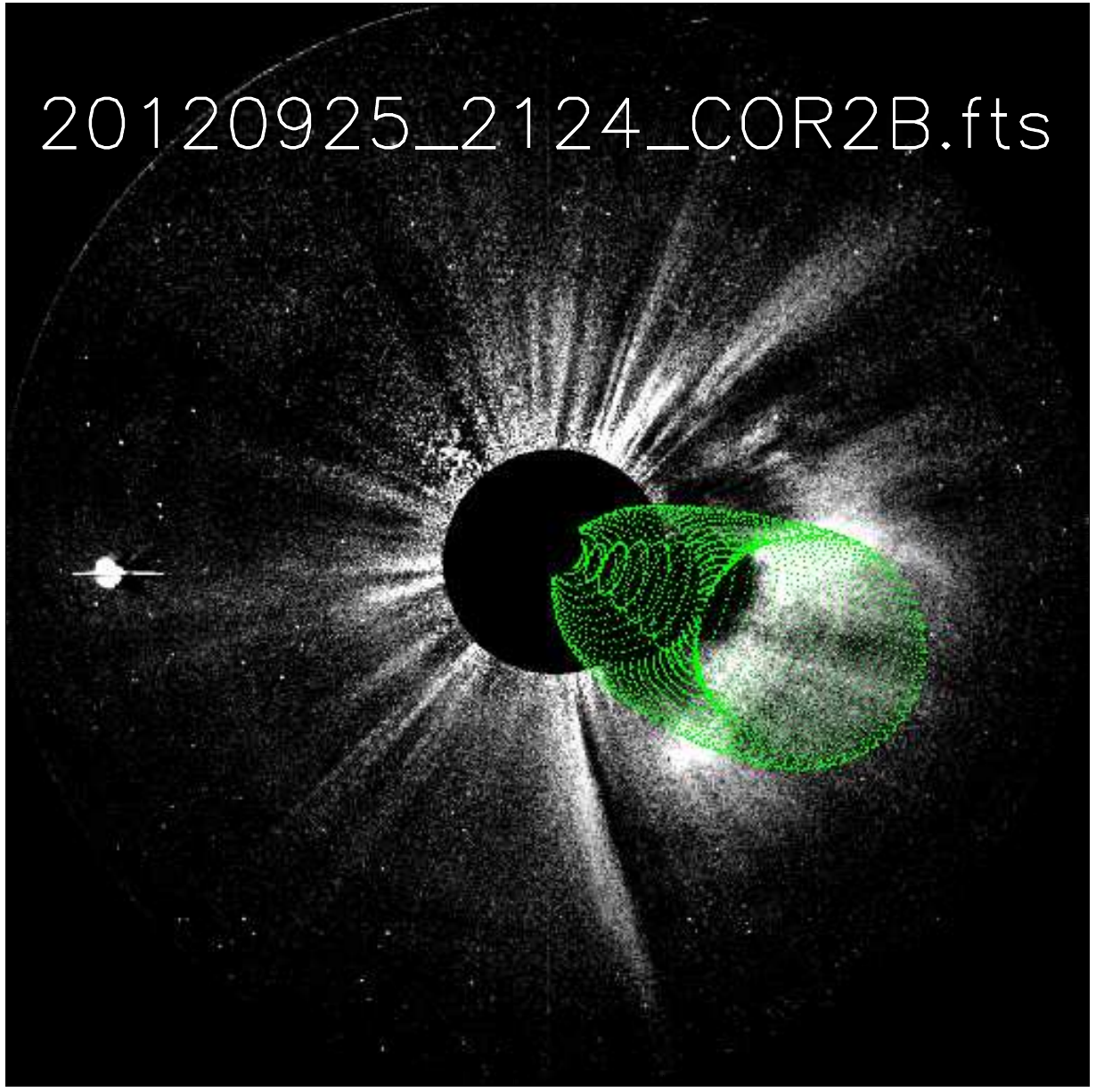}
\includegraphics[angle=0,scale=0.35]{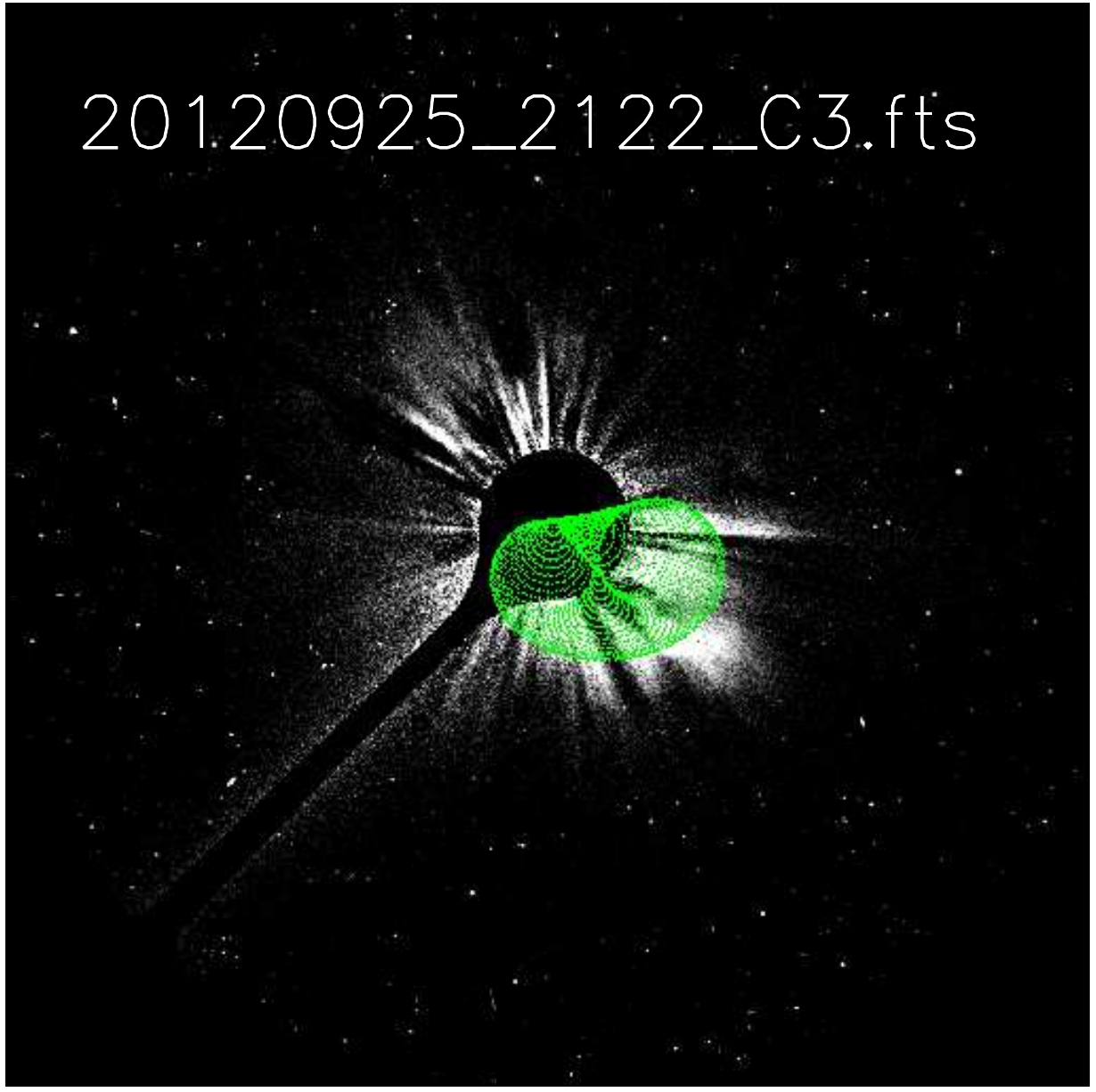}
\includegraphics[angle=0,scale=0.35]{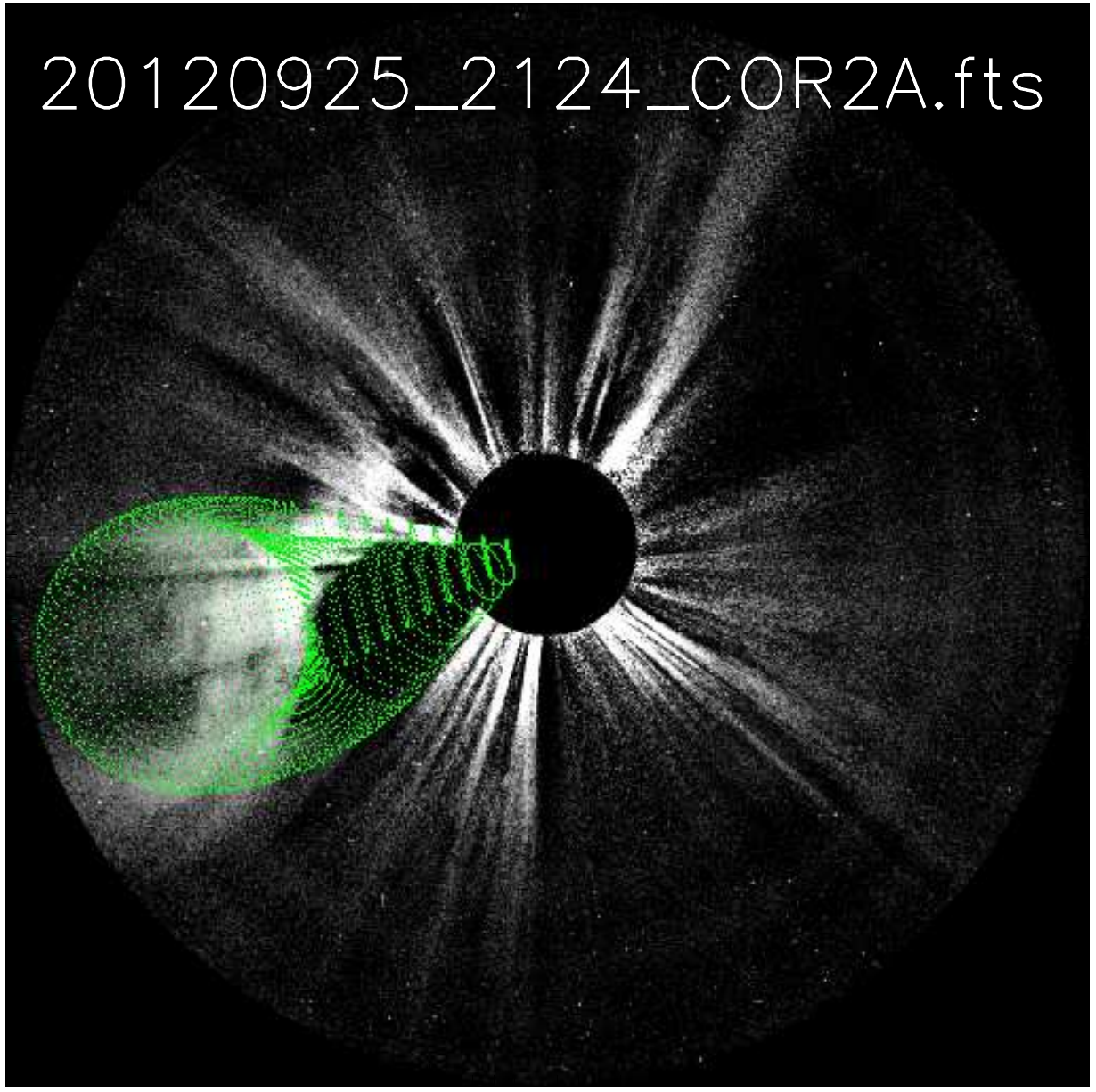}
\includegraphics[angle=0,scale=0.35]{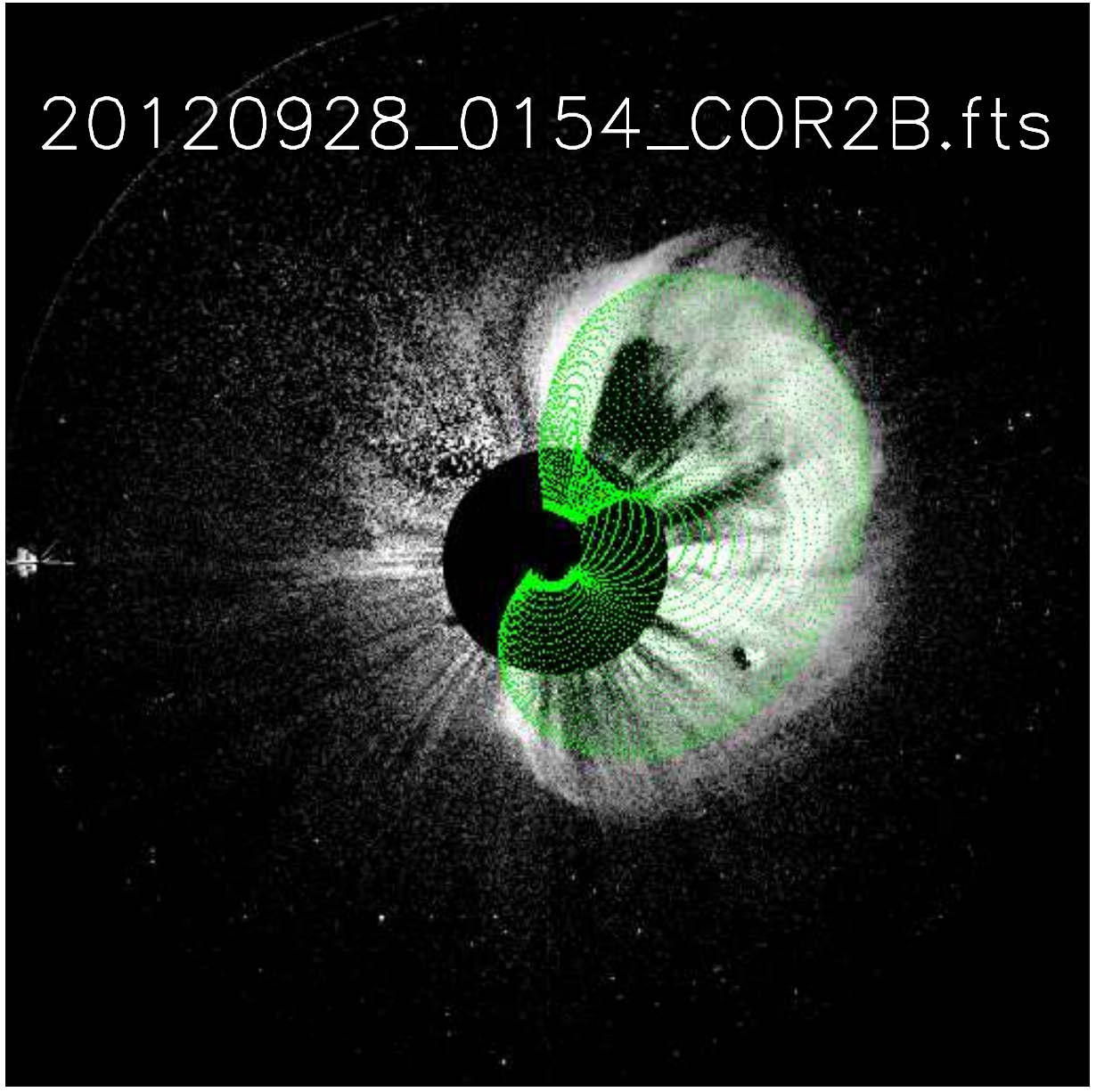}
\includegraphics[angle=0,scale=0.35]{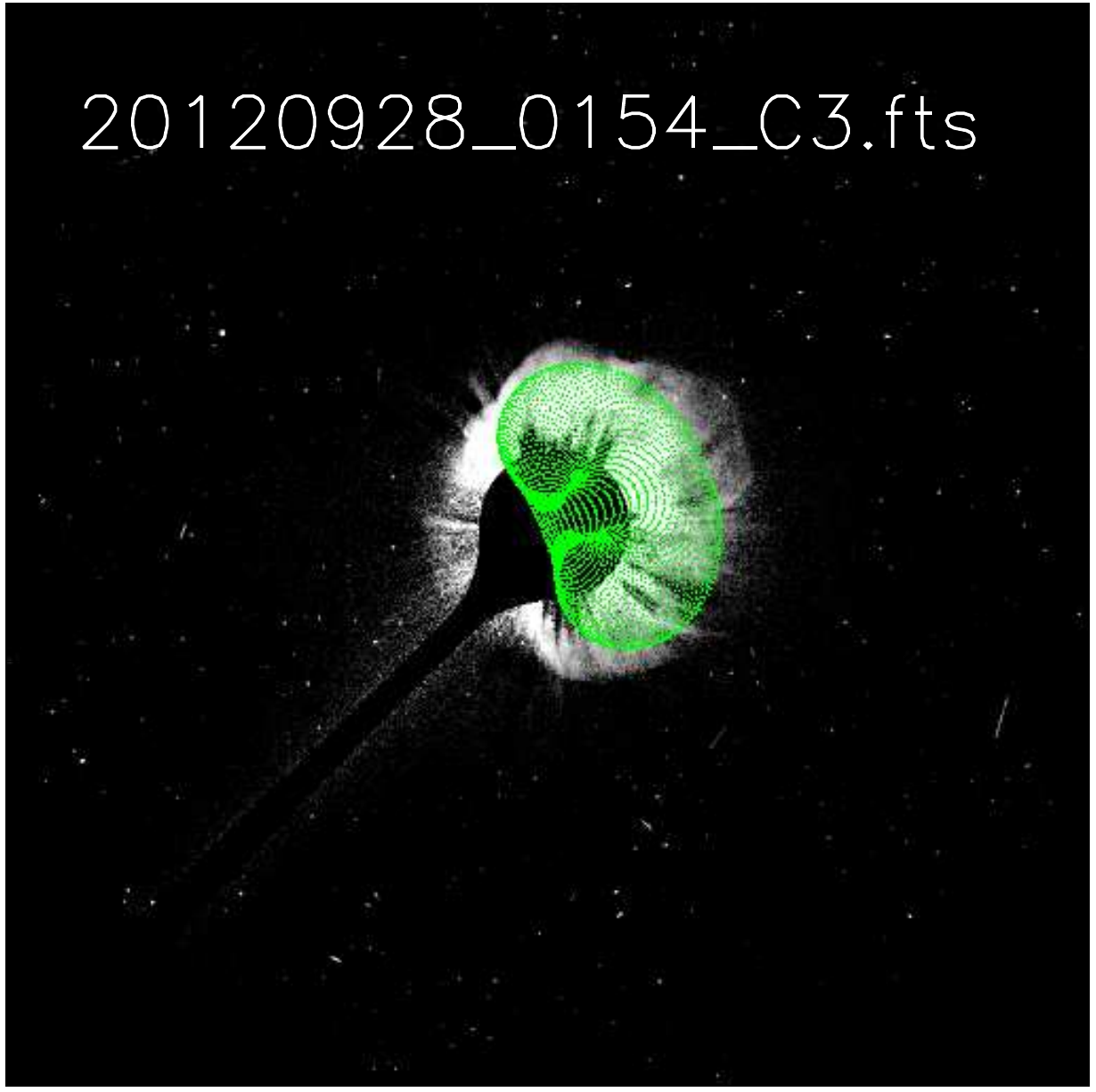}
\includegraphics[angle=0,scale=0.35]{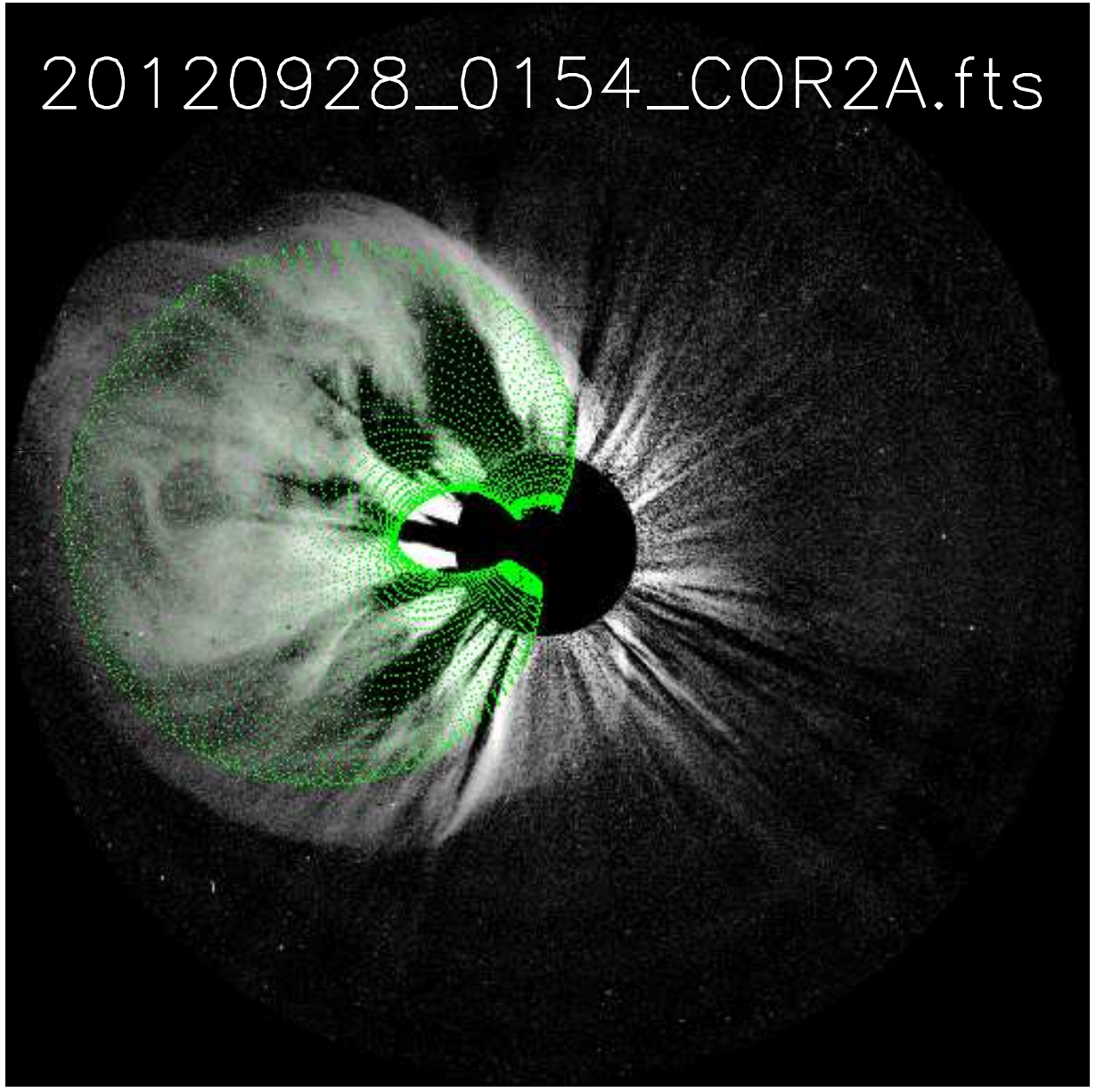}\\
\caption{The GCS model wire-frame overlaid on the CME1 (top panels) and CME2 (bottom panels) respectively. The contemporaneous images triplets are taken from \textit{STEREO}/COR2-B (left), SOHO/LASCO-C3 (middle), and \textit{STEREO}/COR2-A (right) around 21:24 UT on September 25 for CME1 and 01:54 UT on September 28 for CME2, respectively.}
\label{GCS}
\end{center}
\end{figure}

As the \textit{STEREO} spacecraft separation on September 25, 2012 was large, i.e. \textit{STEREO-A} and \textit{B} were approximately $125^{\circ}$ west and $117^{\circ}$ east of the Earth at distances of 0.97 and 1.06 AU from the Sun, respectively, 3D reconstruction for COR2 images is best achieved using the GCS model developed by  \citet{Thernisien2009}. Therefore, in order to estimate the 3D kinematics of CMEs in COR2 field-of-view, we have applied the GCS model   to the simultaneous  images obtained for the CMEs of September 25 and 28, by the SECCHI/COR2 B, SOHO/LASCO-C3 and SECCHI/COR2A coronagraphs.  For this purpose, the processing of total brightness images was carried out followed by the subtraction of a pre-CME image from the sequence of images which were  used for implementing the GCS model \citep{Mierla2008,Srivastava2009,Thernisien2009}. Figure~\ref{GCS} shows images of  September 25 (CME1) and September 28 (CME2) overlaid with the fitted GCS wireframed contour (in green). The best visual fitting of the GCS model  to the CME of September 25 is found in the direction W19S11 (within error of $\pm$ 3$^\circ$) with a half-angle ($\alpha$) of 21$\pm$6$^{\circ}$, a tilt angle ($\gamma$) of  6$\pm$15$^{\circ}$ and an aspect ratio ($\kappa$) of 0.34$\pm$0.10 at a distance of 15$\pm$1.0 R$_\odot$. Visual fitting of the GCS model to the September 28 CME yields its direction of propagation as W25N13 (within error of $\pm$ 3$^\circ$) with $\alpha$ = 68.5$\pm$6$^{\circ}$, $\gamma$ = -75$\pm$15$^{\circ}$ and $\kappa$ = 0.52$\pm$0.10  at a distance of 13$\pm$1.0 R$_\odot$. We carried out several independent attempts of fitting these GCS parameters for the selected CMEs. The aforementioned uncertainties in the fitted parameters are noted by inspecting the differences from each attempt. The longitude values of CME1 \& CME2 as 19$^{\circ}$ and 25$^{\circ}$ suggest that they are propagating westward from the Sun-Earth line. 

We have also applied the Stereoscopic Self-Similar Expansion (SSSE) reconstruction method \citep{Davies2013} on the HI1 images to estimate longitude for CME1 and CME2. In SSSE method, a CME is approximated as a self-similarly expanding sphere not anchored at the Sun and CME cross-sectional angular half-width ($\lambda$) is fixed to an appropriate value.  $\lambda$ measures the curvature  of CME front and it is constrained to vary between 0$^{\circ}$ to 90$^{\circ}$.  The smaller value of $\lambda$ means higher curved front of CME, i.e. its lower radius of curvature. Based on the derived GCS model parameters, the derived $\lambda$ value for CME1 and CME2 is around 41$^{\circ}$ and 31$^{\circ}$, respectively (see details in Appendix~\ref{lmbdwidth}). We used the SSSE method with estimated $\lambda$ value for CME1 and CME2 and found their longitude around 12$^\circ$ and 20$^\circ$, respectively. Using Tangent to a sphere (TAS) method \citep{Lugaz2010.apj}, \citet{Liu2014} have estimated the longitude of CME1 and CME2 as 13$^\circ$ and 25$^\circ$, respectively. Using Geometric triangulation (GT) method \citep{Liu2010} in their study \citep{Liu2014}, they estimated the longitude of CME1 and CME2 as 8$^\circ$  and 14$^\circ$, respectively. These values of longitudes are within 5$^\circ$  to those estimated in the present study. As the TAS and GT methods are the special cases of SSSE method with $\lambda$ = 90$^\circ$  and 0$^\circ$, respectively. Our analysis suggests that near the Sun, the use of different values of $\lambda$ in SSSE method does not result in significant difference in the estimated propagation direction of the CME. Based on the temporal evolution of  the height of the CMEs estimated using the GCS method, their 3D speeds were estimated to be of the order of  500 km s$^{-1}$ (CME1) and 1200 km s$^{-1}$ (CME2). This implies that the following CME2 is much faster than the preceding CME1. Although these CMEs were launched almost 62 hr apart,  their HI observations suggest that part of both the CMEs along the same latitude will interact, as they both propagate westward of the Sun-Earth line along the longitudes separated by almost 10$^\circ$ to each other.

\subsection{3D reconstruction of CMEs in HI field-of-view}
\begin{figure}[!htb]
\captionsetup{width=0.85\textwidth}
\begin{center}
\includegraphics[angle=0,scale=0.59]{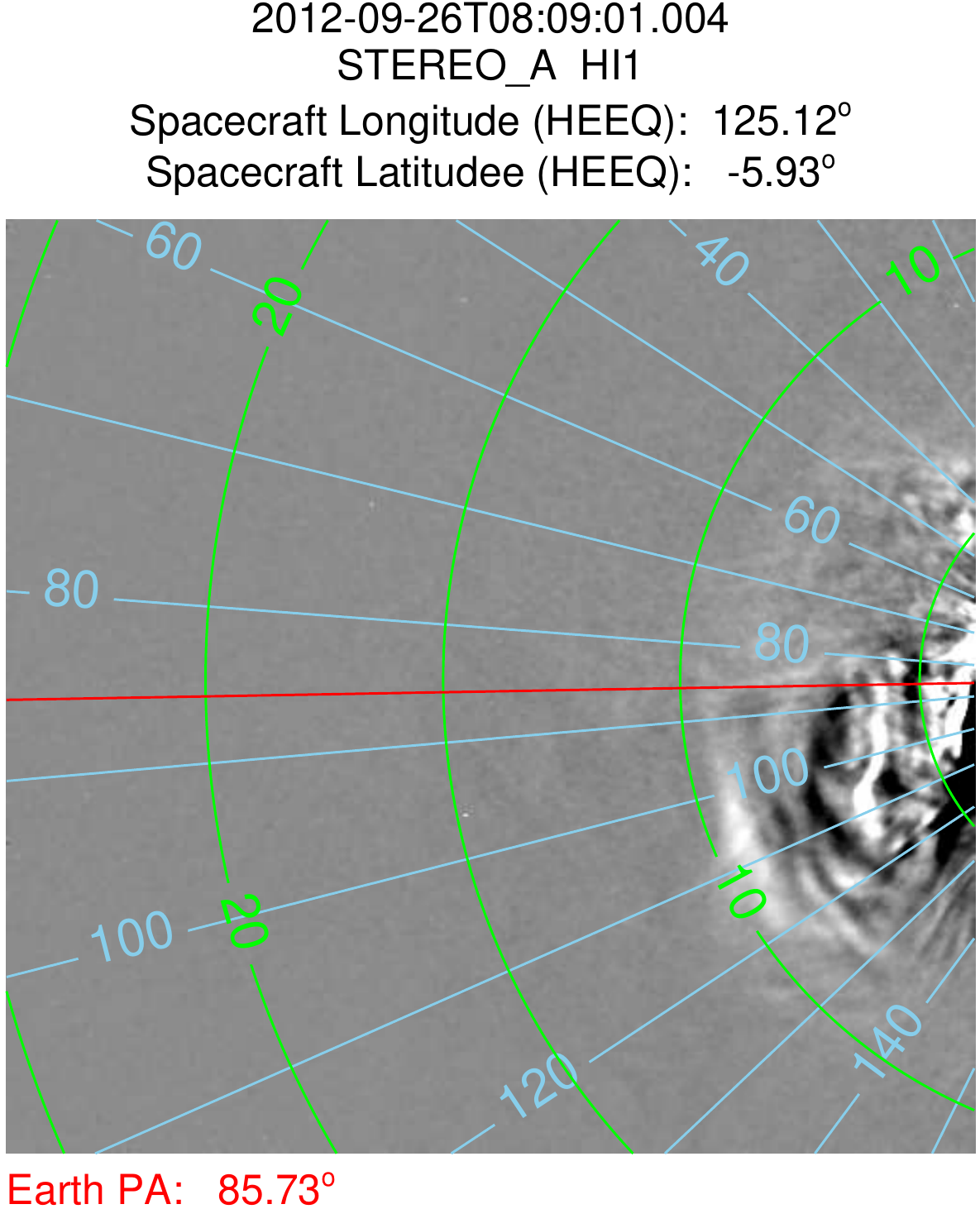}
\includegraphics[angle=0,scale=0.59]{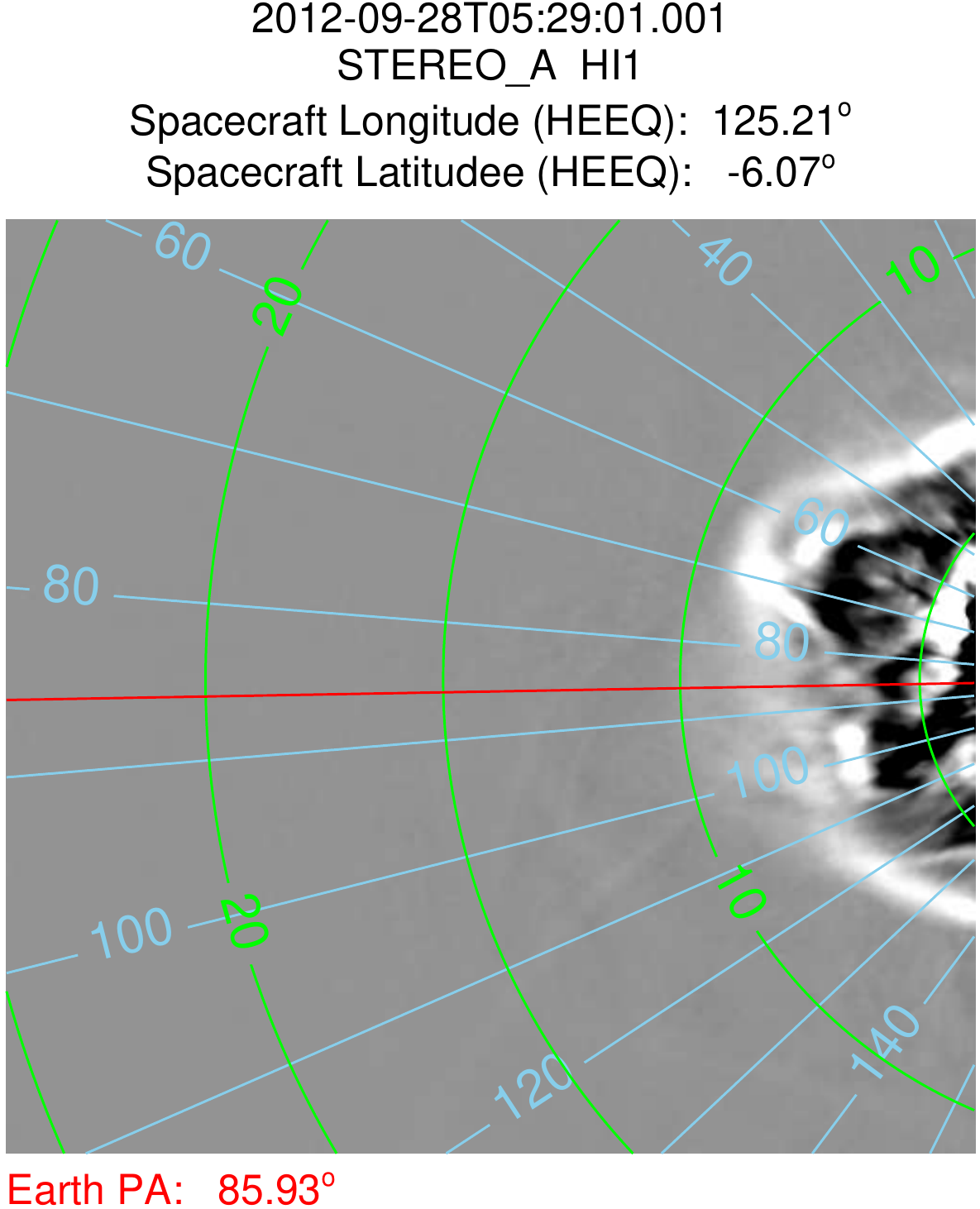}
\includegraphics[angle=0,scale=0.59]{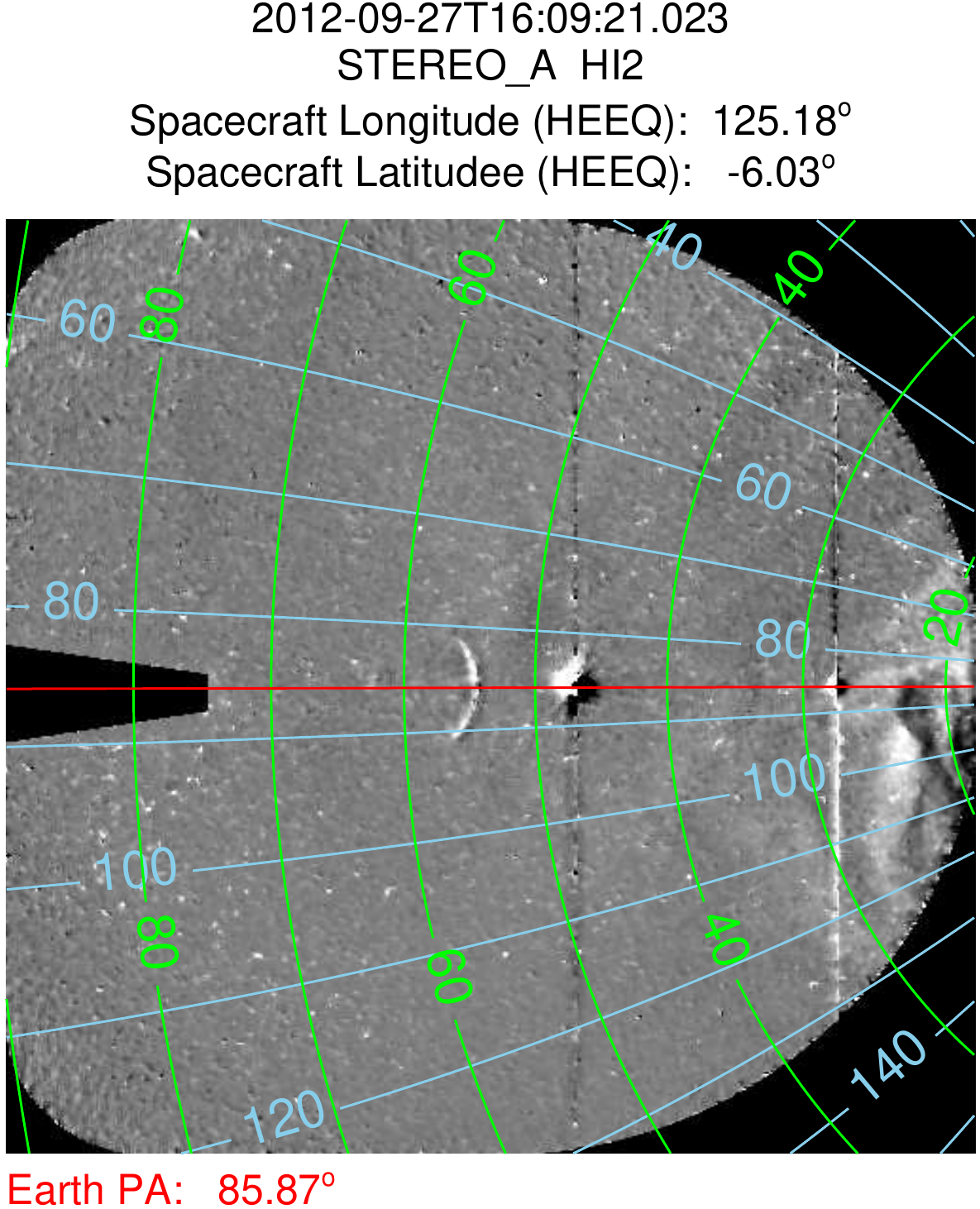}
\includegraphics[angle=0,scale=0.59]{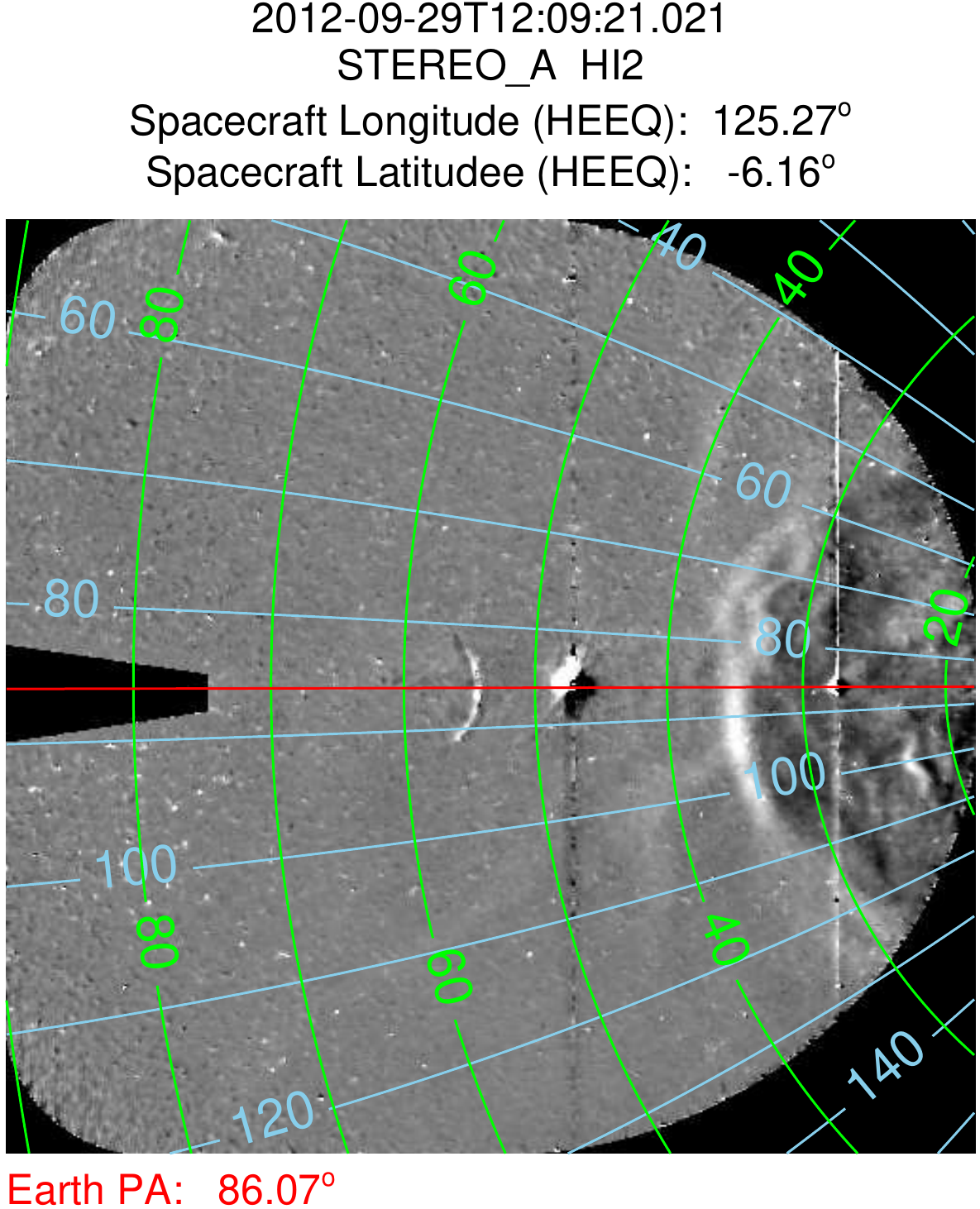}
\caption{Evolution of the CMEs of September 25 (left panel) and September 28, 2012 (right panel) in running difference images of the HI1-A (top panel) and HI2-A (bottom panel)  obtained from \textit{STEREO}/SECCHI Ahead spacecraft.}
\label{Evolution}
\end{center}
\end{figure}

\begin{figure}[!htb]
\captionsetup{width=0.85\textwidth}
\begin{center}
\includegraphics[angle=-270,scale=0.50]{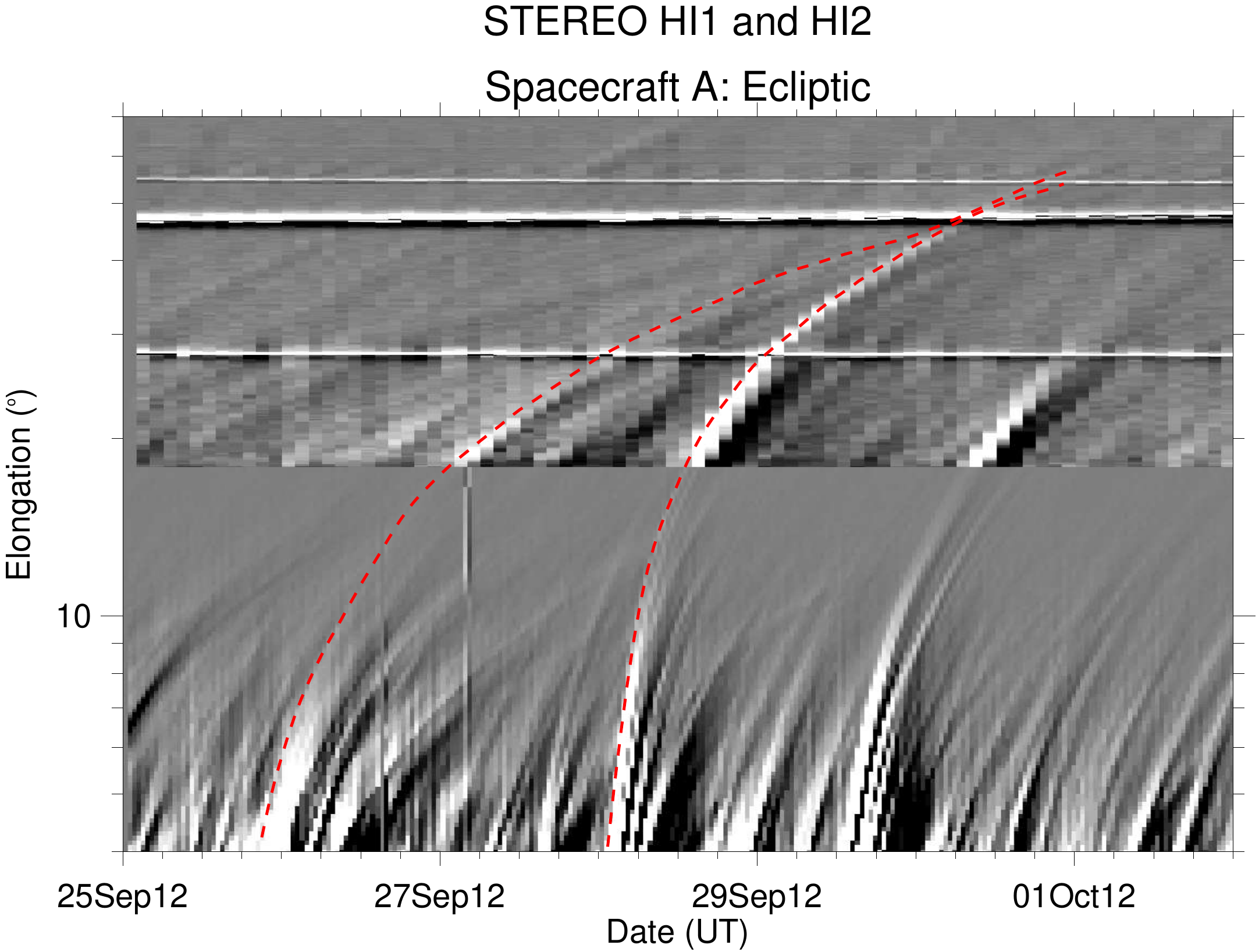}
\caption{Time-elongation plot or J-map constructed using HI1-A and HI2-A images of \textit{STEREO}/SECCHI Ahead spacecraft for the period of September 25-October 2, 2012. The two red tracks (dashed line) mark the outline of the leading edges observed as brightness enhancement for the two CMEs.}
\label{Jmap}
\end{center}
\end{figure}

As described above, the estimated direction and speed of both the CMEs suggest their probable interaction far out in the heliosphere. Therefore, we tracked these CMEs further in HI images and estimated their true kinematics. For this purpose, the HI1 and HI2 \citep{Eyles2009} images were obtained from the Level 2 data-set of the UKSSDC (http://www.ukssdc.ac.uk/solar/stereo/data.html). We  then subtracted a long-term background image from the sequence of images that were used for estimating the kinematics of CMEs. After subtracting the background, the stellar contribution was also removed by aligning the HI image pair before taking their running difference. Then the obtained difference images in HI field-of-view were  examined carefully to track any density depletion or enhancement due to the  passage of  the CMEs. The evolution of these CMEs in HI1-A and HI2-A field-of-view is shown in Figure~\ref{Evolution}.  From the figure, it is clear that the heliospheric tracking of a CME feature out to longer elongation is difficult. Therefore, J-map is constructed using a sequence of running difference images by extracting the brightness profile as a function of elongation, averaged over a narrow range of PA along the ecliptic, and stacking them vertically as a function of time on the X-axis \citep{Sheeley1999,Davies2009}. The detail of the construction of the J-map and derivation of elongation of moving CME feature is described in  \citet{Mishra2013}. The J-map of the selected CMEs for HI-A observations along the ecliptic is shown in Figure~\ref{Jmap}.  The two tracks (marked in red) on the J-map reveal the time evolution of the brightness enhanced features corresponding to the leading edges of the CMEs of September 25 and 28, 2012 (Figure~\ref{Jmap}).

It is to be noted that as the J-maps are constructed along the ecliptic, therefore only a portion of the CME moving in the ecliptic plane is tracked in our study. Although the selected CMEs are propagating along the different latitudinal direction separated by around 37$^\circ$, but our approach (J-map and SSE method) estimates their speed along the ecliptic plane (i.e. around 0$^\circ$ latitude). This estimated speed is equal to the projection of radial speed (in direction of CME latitude) onto the ecliptic plane. The difference in the longitude direction of both the CMEs is around 6$^\circ$ which is small in contrast to the angular size of the CME. Therefore, we have considered that the tracked portion of both the CMEs are moving along the same direction in the heliosphere. We notice in Figure~\ref{Jmap} that the tracks corresponding to both the CMEs are approaching each other and meeting at 48$^{\circ}$ elongation. This is an indication of possible collision between both the Earth-directed CMEs around 6:00 UT on September 30.

Stereoscopic reconstruction methods estimate more accurate kinematics (distance and speed) than single spacecraft reconstruction methods and additionally  provide continuous time-variations of the CME direction \citep{Mishra2014}. The selected CMEs are not well observed in the HI-B (specially HI2) images and their intensities even in J-map, constructed using HI-B images, are too weak to track them unambiguously  beyond 20$^{\circ}$. These CMEs are well observed only in HI-A field-of-view and therefore we could not implement a stereoscopic reconstruction method to estimate the CME direction near the collision site. Instead, we used the single spacecraft reconstruction method, namely the  SSE method \citep{Davies2012}. The use of SSE method allows to assume a circular cross-section of the CME with a certain angular width, in the plane corresponding to the position angle of interest. The SSE geometry can be characterized by $\lambda$ value and in its extreme forms with $\lambda$ equal to 0$^{\circ}$ and 90$^{\circ}$, it is equivalent to other single spacecraft reconstruction methods like the Fixed-Phi (FP) method \citep{Kahler2007} and  the Harmonic Mean (HM) method \citep{Lugaz2009}, respectively.  

\begin{figure}[!htb]
\captionsetup{width=0.85\textwidth}
\begin{center}
\includegraphics[angle=0,scale=0.90]{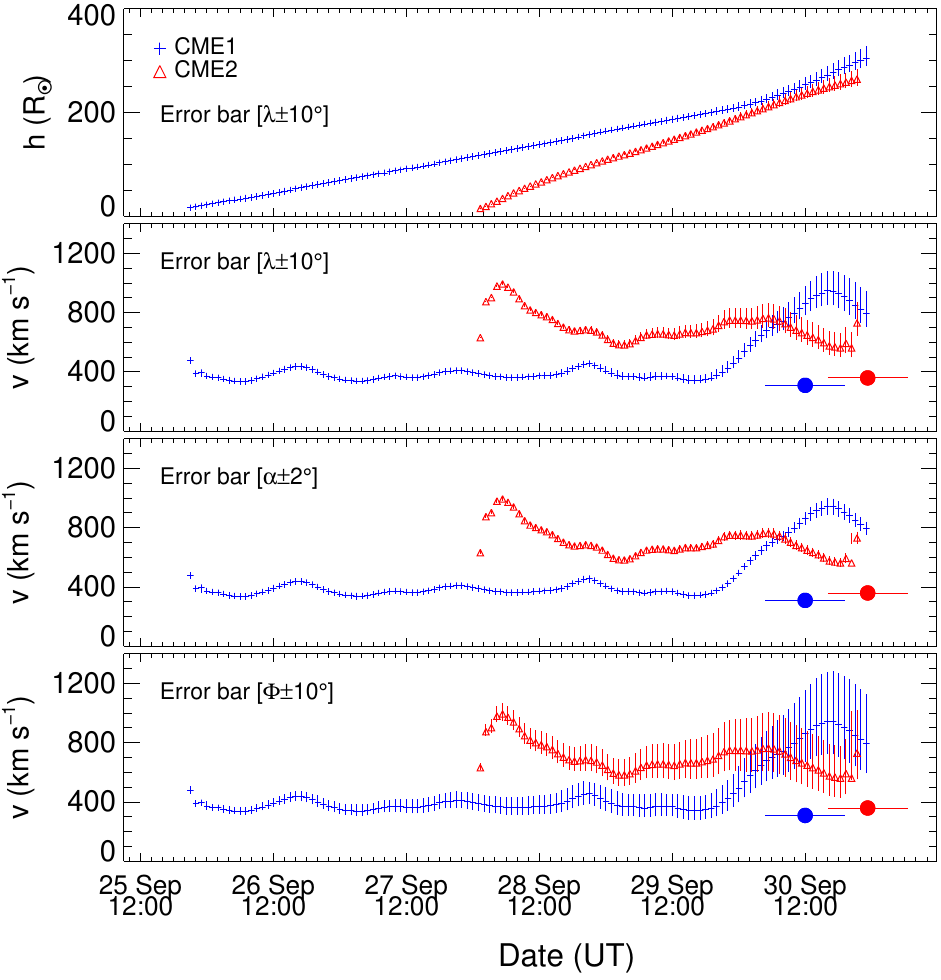}
\caption{3D kinematics of the interacting CMEs of September 25 and 28, 2012 using the SSE method. The vertical lines show the errors bars. In the top two panels, errors are calculated using $\lambda$ that are +10$^\circ$ and -10$^\circ$ different to the $\lambda$ value of the CMEs derived using the GCS model parameters.  The errors shown in third and fourth panels from the top, respectively correspond to effect of uncertainties of $\pm$2$^{\circ}$ in observed elongation ($\alpha$) and $\pm$10$^{\circ}$ in observed propagation direction ($\Phi$) of the CMEs. The horizontal lines and filled circles in the speed panels, respectively represent the in situ measured speed and arrival times of the CMEs.}
\label{HT_lamerr}
\end{center}
\end{figure}

We implemented the SSE method for CME1 and CME2 using their $\lambda$ values as 41$^{\circ}$ and 31$^{\circ}$, respectively as estimated from GCS model parameters and described in Section~\ref{3DCOR}. In SSE method, it is assumed that CME propagation direction, estimated using GCS model in COR field-of-view, is constant in the heliosphere. Although direction estimates in HI field-of-view using SSSE method can account for CME deflection, we preferentially rely on the direction estimates from GCS model. This is because unlike the SSSE, GCS model has been used on an additional viewpoint (i.e. SOHO) observations of the CMEs.  The effect of changes in the values of direction on the kinematics from SSE method, is analyzed in in Section~\ref{KinInt}. The estimated kinematics from SSE method is shown in Figure~\ref{HT_lamerr}.  In the speed panels, the horizontal lines and filled circles respectively marks the in situ measured speed and arrival time of both the CMEs. This shows the inconsistency in the speed measured from remote and in situ observations. It is clear that even before the noticeable merging of CMEs height-time tracks (Figure~\ref{HT_lamerr}), CME1 gets unphysical acceleration. This led us to examine the uncertainties in kinematics because of the errors in assumed CME propagation direction, $\lambda$ value and in elongation angle.  Hence, the use of such unphysical kinematics can lead to large errors in the study of nature of collision for these interacting CMEs.

Using uncertainties in GCS fitted parameters, we estimated an uncertainty of $\approx$ $\pm$ 10$^{\circ}$ in the $\lambda$ values and examined its effect on the estimated kinematics. The error bars shown in top two panels of Figure~\ref{HT_lamerr} represent the uncertainties in the kinematics because of using $\lambda$ value that are +10$^\circ$ and -10$^\circ$ different to the values calculated based on GCS model parameters. We consider an uncertainty of $\pm$2$^\circ$ in elongation angle measurements and its effects on the kinematics is shown as error bars in the third panel (from the top) of  Figure~\ref{HT_lamerr}. We have also considered the effect of $\pm$10$^{\circ}$ change in propagation direction of the CMEs on their estimated kinematics. This is shown as error bars in the bottom panel of  Figure~\ref{HT_lamerr}. 

\subsection{Examining the effect of $\lambda$ on SSE method}
\label{SSEdiflmbd}
The selected Earth-directed CMEs in our study are far-sided for \textit{STEREO-A} observer  located around 105$^{\circ}$ away from  propagation direction of the CMEs. The effect of varying $\lambda$ on the kinematics estimated from SSE method, is crucial to examine for such CMEs whose nose can never be seen by the observer. Figure~\ref{loca} illustrates the effect of position of observer on the estimated height of the CME leading edge.  See Appendix~\ref{lmbdSSE} for details about this effect. From the figure, it is clear that use of different value of $\lambda$ in SSE method lead to different estimates of distance, especially for the far-sided CME \citep{Liu2013}.

Although both the selected CMEs in our study have different values of 3D width (estimated from GCS model) but their value of $\lambda$ remains almost the same (around 30-40$^\circ$) because of their extremely different orientation of flux-rope. Therefore, assuming $\lambda$ to be equal for both the CMEs, we exploited the capability of the SSE  method to vary the angular width of the CMEs and estimated their 3D kinematics. By  varying the values of the angular width of CMEs to 0$^{\circ}$, 30$^{\circ}$, 90$^{\circ}$, we examined the variations in the kinematic profiles of these interacting CMEs which are far-sided for the observer (i.e. \textit{STEREO-A}). Figure~\ref{HT_difflambda} shows the time evolution of height and speed of the leading edges of the two CMEs corresponding to different chosen value of $\lambda$.  The impact of these variations have been examined in details which  are discussed in the next section. In the speed panels of the figure, the in-situ measured speed and arrival times of the CMEs are shown respectively with solid horizontal lines and filled circles. Such analysis may help to better understand, interpret the HI observations and also the effect of geometry on the CMEs which are back-sided for \textit{STEREO}

\section{Results and Discussion}
We examine the kinematics of the interacting CMEs in the heliosphere estimated using SSE method with different angular width for the CMEs. Based on the estimated kinematics, the signatures of interaction and the nature of collision is presented in Section~\ref{NatCol}. The in-situ signatures of the interacting CMEs and their role in producing a strong geomagnetic storm is summarized in Section~\ref{Insitu}.

\subsection{Kinematics of the CMEs}
\label{KinInt}
From the variation of 3D height with time and assuming different angular widths ($\lambda$ = 0$^\circ$, 30$^\circ$, 60$^\circ$) for the two CMEs (as shown in Figure~\ref{HT_difflambda}), we note the following:

Firstly, when $\lambda$ = 0, the CME1 shows unphysical acceleration at a distance close to the Earth, therefore we interpret that our assumption of $\lambda =0$ does not hold good. The speed derived with lower value of $\lambda$ is highly inconsistent with in situ measured speed of the CME, even if its flank is intercepted by the in situ spacecraft. Such unphysical acceleration of the CMEs may be due to the erroneous fixing of their longitude in SSE method in case of a real deflection \citep{Webb2009,Wood2010,Mishra2014} or an effect of expanding CME geometry \citep{Howard2009a,Howard2011}. Secondly, as we consider 
$\lambda$ = 30$^\circ$ and 90$^\circ$, the unphysical acceleration reduces noticeably. This is explained in Appendix~\ref{lmbdacc}. 

We note that SSE method with  $\lambda$=90$^{\circ}$  (i.e. HM method) serves the best to represent the kinematics of the selected interacting CMEs and therefore is considered for studying the  characteristics of collision. However, if a CME spherical front is flattened by its interaction with structured solar wind or a CME is narrow, then HM method may be less accurate \citep{Wood2009}. In general, the effect of flattening also partially contributes in overestimation of CME distance and speed. Under this effect, higher value of $\lambda$, i.e. less curved CME front in SSE method will suffer lesser with overestimation issue. Such finding that HM method \citep{Lugaz2009} performs the best amongst the single spacecraft methods is also  noted in \citet{Mishra2014}.  Our result also emphasizes that one needs to be careful in choosing the angular width of an Earth-directed CME for estimating the 3D kinematics using single spacecraft methods, in particular when each \textit{STEREO} spacecraft has separation of more than 90$^{\circ}$ from the Sun-Earth line, during February 2011 to June 2019. 

Another point to be noted from this analysis, specially in the scenario when the \textit{STEREO} observe back-sided CMEs,  is that close to the Sun the kinematics of the CMEs do not vary significantly on  changing the angular width of the CMEs. They become significant only as a CME propagates and expands further out in the heliosphere. However, when the \textit{STEREO} spacecraft are positioned in front of the sun, relatively small difference in the estimated kinematics for an Earth-directed CME (for different values of angular width of the CMEs) is noticed \citep{Wood2010,Howard2012b,Mishra2014}. A more detailed work is required to examine the least and most noticeable effect of observer position for different possible geometry for the CME launched in different direction. We also emphasize that the above findings are based on a crucial assumption that the longitude does not change for the selected CME during its propagation in the heliosphere.  Any change in the longitude which is likely during collision, can lead to erroneous estimation of the CME speed from SSE method.  However, the sudden unphysical increase in the speed corresponding to smaller value of $\lambda$ is noticed before the collision, and is mostly because of incorrect use of $\lambda$ rather than $\Phi$ of the CMEs.

We examined the uncertainty in the kinematics which owes from uncertainty in propagation direction used in SSE method. For this, we have repeated our SSE analyses using propagation directions of the CMEs that are +10$^\circ$ and -10$^\circ$ different than the values ($\Phi$) estimated using GCS model. The resulting uncertainties in the distance and speed, for different values of $\lambda$, are shown with vertical lines on each data points in top six panels of Figure~\ref{HT_difflambda}. The lower and upper segments of the vertical lines are uncertainties corresponding to the $\Phi$+10$^\circ$  and $\Phi$-10$^\circ$, respectively.  We find that when the CME is propagating in a direction $\Phi$-10$^\circ$, i.e. farther from the observer, its distance and speed are estimated larger than those from $\Phi$. The effects are opposite for the CME propagating closer to the observer than that of $\Phi$. This is true, irrespective of the values of $\lambda$ considered in SSE method, however crucial for smaller values of $\lambda$. We have also considered a reasonable uncertainties of $\pm$2$^{\circ}$ in elongation angle measurements and examined its effect on the estimated kinematics of both the CMEs. This is shown with vertical lines at each data points in the bottom three panels of Figure~\ref{HT_difflambda}. We note that a reasonable error in elongation angle measurements has a smaller effect on the kinematics than the reasonable error in propagation direction. An interesting point to note is that for equal uncertainties in the propagation direction of $\pm$ 10$^\circ$, the magnitude of change in the speed is not the same. Figures 3, 6 and 8 of  \citet{Mishra2014} show almost an equal change in the speed because of $\Phi$$\pm$ 10$^{\circ}$ for CMEs directed towards the observer. The error in assumed propagation direction ($\Phi$) of the CMEs has significant effect on the kinematics at larger distance from the Sun, specially for smaller value of $\lambda$. Hence the estimates of kinematics for a far-sided CME have huge errors at higher elongation.

\subsection{Interaction phase and nature of collision}
\label{NatCol}

As explained earlier about the limitations of the methods at a large  distance from the Sun, i.e. close to the Earth where collision took place, we rely on the pre and post collision kinematics estimated with $\lambda$ = 90$^ \circ$ for understanding the collision characteristics of  interacting  CMEs of September 25 and 28, 2012. The collision phase is marked by two vertical dashed lines in Figure~\ref{HT_difflambda}, in the third and sixth panel from the top. As discussed in the earlier papers \citep {Mishra2014a,Mishra2015} the start of the collision phase is defined by the instant when the speed of the CME1 starts to rise with a simultaneous decrease in the speed of CME2. The end of the collision is defined by the time when either the two CMEs attain an approximately equal speed or their trend of acceleration is reversed. In the present case, for $\lambda$ = 90$^\circ$, the duration of collision phase is 16.8 hr which starts on  September 29 at 22:48 UT and ends on September 30 at 15:35 UT. From the plot, we find that at the beginning of the collision phase, the initial speeds of CME1 and CME2 are 385 km s$^{-1}$ and 610 km s$^{-1}$ respectively. At the end of the collision phase the values of speeds of CME1 and CME2 are found to be 710 km s$^{-1}$ and 430 km s$^{-1}$ respectively. The true masses of the two CMEs were also computed from the COR2 images, following the method of \citet{Colaninno2009} and were found to be $1.75 \times 10^{12}$ kg and $9.67 \times 10^{12}$ kg for the CME1 and CME2 respectively. To calculate the coefficient of restitution ($e$) of the colliding CMEs,  we have constrained the conservation of momentum to be valid as a necessary condition for the collision. We also assumed no change in the estimated mass of the CMEs beyond COR2 field-of-view. The procedure and equations used for calculating the value of $e$ is described in \citet{Mishra2015, Mishra2014a}.  

Precisely, our approach calculates a set of  theoretical values of post-collision speeds (v$_{1th}$, v$_{2th}$) closest to the observed  post-collision speed (v$_{1}$, v$_{2}$) by varying the values of $e$ such that they satisfy conservation of momentum with observed mass and pre-collision speed of the CMEs. For this theoretically calculated speed, the variance  $\sigma = \sqrt{(v_{1th} - v_{1})^{2} + (v_{2th} - v_{2})^{2}}$ is minimum corresponding to obtained value of $e$. Using the approach, $e$ is estimated to be 0.8.  For this value of $e$, the final speed of the CMEs after their collision is calculated as  (v$_{1th}$, v$_{2th}$) = (728,548) and $\sigma$ = 120 is noted.  The total kinetic energy before the collision was 1.92 $\times$ 10$^{24}$ J, which decreased by $\approx$ 0.6\%  for the derived value of  $e$ =0.8,  for the collision. Due to this, momentum of CME1 increased by 89\% while that of CME2 decreased by 10\%. The kinetic energy of CME1 and CME2 before the collision was 1.29 $\times$ 10$^{23}$ J and 1.79 $\times$ 10$^{24}$ J, respectively. Based on the post-collision speed for the estimated value of $e$, we found that after the collision, the kinetic energy of CME1 increased by 258\% to its value before the collision, while the kinetic energy of CME2 decreased by 19\% to its value before the collision. Such exchange in the momentum and kinetic energy of the CMEs is expected as CME2 is 5.5 times massive than preceding CME1. However, if we directly estimate the value of $e$ from the pre- and post-collision speed of CMEs without considering the conservation of momentum, the value of $e$ is determined as 1.2. This is because of the errors in the speed estimated from the reconstruction method, and therefore a different approach is undertaken in our study to derive the value of $e$ by constraining the conservation of momentum.

We acknowledge that our approach to use speed of leading edge, which consists of the propagation speed and expansion speed, may lead to some errors in the estimated coefficient of restitution. However, the errors in the leading edge speed may arise due to several sources and their accurate quantification is extremely difficult. Considering huge uncertainties (and fluctuations) in the estimated leading edge speed of the CME in our study and acknowledging the possible errors in estimating the expansion speed of the CME in HI field-of-view, we have not carried out the analysis taking separately both the speeds  into account. In our case, the collision takes place almost near the Earth and at such large distance from the Sun, the expansion speed of the CME is relatively smaller than the leading edge speed \citep{Poomvises2010}. We consider that error in our analysis comes mainly from error in the estimated speed of the CMEs  and therefore we have repeated the calculation of $e$ by taking an uncertainty of $\pm$ 100 km s$^ {-1}$ in their initial and final speeds.  

The simultaneous increase or decrease of 100 km s$^ {-1}$ in estimated initial and final speed  of CMEs, the $e$ values is calculated as 0.8 with $\sigma$ = 120. If the final (initial) speeds of the CMEs are considered to be smaller (larger) by 100 km s$^ {-1}$ with no change in their initial (final) speeds, the value of $e$ becomes as 0.4 corresponding to $\sigma$ =235. If the final (initial) speeds of the CMEs are considered to be larger (smaller) by 100 km s$^ {-1}$ with no change in their initial (final) speeds, the value $e$ becomes as 1.25 with $\sigma$ = 5. If initial speeds of the CMEs are increased and their final speeds are decreased, $e$ becomes equal to 0 with $\sigma$ = 351. An increase in the final speeds and decrease in the initial speeds results in acceleration of both the CMEs and does not satisfy the collision scenario. The higher values of $\sigma$ means that theoretically calculated final speeds are highly inconsistent with the observed final speeds considered in our approach. As the CMEs speeds from SSE methods are already overestimated (see Section~\ref{Insitu}), a further uncertainty of +100 km s$^ {-1}$ in their speeds is rarely possible. Therefore, based on careful inspection of $\sigma$ values, we note reliable calculated value of $e$ varies from 0.8 to 1.2. Hence, the values of $e$ = 0.0 or 0.4 are meaningless and can not be considered as representative of collision nature for the CMEs. Such error analysis emphasizes the involved inaccuracy  in the estimated value of $e$ due to uncertainties in the derived speeds of CMEs from HI based observations. We also note a huge error in the speed of the CMEs during the collision phase as shown in Figure~ \ref{HT_difflambda}, and this makes the estimated value of $e$ somewhat less reliable. Further, before the actual collision of the CMEs, CME2 driven shock has contributed in partial acceleration of CME1 (see Section~\ref{Insitu}), and also CME2 might have decelerated because of its magnetic interaction with CME1. Therefore, marking of collision phase also leads to uncertainties in the speeds of the CMEs.  The true mass of the CME estimated in COR field-of-view may vary as the CME propagate and even if it does not, it is difficult to know whether the total mass of the CME participates in the collision process.  Acknowledging these assumptions taken in the present study,  we find that the nature of collision for the selected CMEs was close to elastic in nature. Under the light of previous studies \citep{Shen2012,Temmer2014,Mishra2014a,Mishra2015}, it is imperative to find what causes the nature of collision of CMEs to lie in a broad range from super-elastic to perfectly inelastic.

The tracked features of selected CMEs are moving along the longitude within 6$^{\circ}$ to each other in the ecliptic plane, hence our approximation of one-dimensional i.e. head-on collision scenario is reasonable. However, if the collision changes the direction of the participating CMEs, then it must be taken into account for calculating the modified speed. It has been shown earlier that approaching direction of two colliding CMEs has a significant influence on the collision nature \citep{Shen2012}. As there are only few points of elongation-time curve in HI2-A,  and therefore any fitting methods \citep{Rouillard2008,Lugaz2010,Davies2012}) can not be reliably applied to estimate the post-collision direction of propagation of the CMEs \citep{Williams2009}. We rely on the direction of CMEs estimated in the COR field-of-view and consider it unchanged during the entire heliospheric journey of the CMEs. We note that error in speed measurements itself lead to significant error in the calculation of $e$ values, even if the other possible sources of errors are neglected. 
      
\subsection{In-situ observations of interacting CMEs}
\label{Insitu}

We analyzed the in situ measured plasma and magnetic field observations at 1 AU and identified the arrival of the CME1 and CME2. The variations in plasma and magnetic field parameters from 2012 September 30 to October 3 are shown in Figure~\ref{insitu}.  Our findings on identification of CMEs and their signatures of interaction, from in situ data analysis, are in agreement to a large extent with that reported by \citet{Liu2014}. However, for sake of completeness, we summarize the important points. Moreover, we have also made an attempt to identify the filament in in situ observations and results of CME interaction is compared to our earlier analysis in \citet{Mishra2014a} and \citet{Mishra2015}. 

For instance, in the present case the CMEs arrive at the Earth as two distinct structures with their associated shocks at 11:55 UT and 23:10 UT on September 30. These shocks are marked as S1 and S2 at the top  in  Figure~\ref{insitu}. The shock driven by CME2 have entered in the CME1 structure, however it has not traversed through  CME1 completely. In the figure, the region bounded between the second and the third vertical lines during September 30, 23:10 UT and October 1, 04:50 UT  shows strong southward component of magnetic field reaching up to -20 nT and enhanced temperature up to 10$^5$ K. This is possibly because of compression of CME1 trailing edge by CME2 driven shock. The region marked as CME1 also does not show signature of expansion and hence seems to be compressed due to force exerted by CME2. Based on the speed, magnetic field strength and  its rotation, CME2 is identified. The structure corresponding to CME2 shows enhanced temperature of the order of 10$^5$ K despite having the signature of expansion. This is not typically found for isolated CME and is the result of collision of CME2 with CME1. However, the presence of plasma material of CME1 trailing edge after the marked leading edge of CME2 cannot be denied, as the merging of both CMEs is in progress and continues beyond the Earth \citep{Liu2014}.  The possibility of such mixing of CMEs have also been reported earlier which emphasize that signatures of CMEs interaction observed in situ depend on the relative orientation of the flux-rope associated with both CMEs  \citep{Xiong2006,Xiong2009,Lugaz2013,Lugaz2014}.

An important point to note from Figures~\ref{HT_difflambda} and ~\ref{insitu} is the disagreement between the CMEs speed from remote and in situ observations. The in situ measured speed of CME1 and CME2 is 310 and 360 km s$^{-1}$, respectively while their speed at L1 using SSE method corresponding to $\lambda$ = 90$^{\circ}$, is  700 and 450 km s$^{-1}$.  From the estimated kinematics using remote observations, we note that arrival time of CME1 and CME2 at L1 is 5 hr and 9 hr earlier, respectively, than their actual arrival time measured at in situ \textit{WIND} spacecraft. Despite the continuous tracking of the CMEs up to L1 and beyond, the error noted in the arrival time is suggestive of overestimated value of the CME speed from SSE method.  Figure~\ref{HT_difflambda} shows that extent of overestimation of speed is reciprocal to the chosen value of $\lambda$ for the CMEs. Also, the description in Appendix~\ref{lmbdacc} and Figure~\ref{HT_difflambda} indicate that speed of the CMEs is overestimated as they are moving away  (i.e. $\Phi$$>$90$^{\circ}$) from the observer. The short interval of CME1 in Figure~\ref{insitu} and the estimated direction of CMEs from GCS and SSSE methods suggest for their flank encounter with in situ spacecraft. The inconsistency in speed from remote and in situ, may be partially due to the fact that remotely tracked feature of the CME has not been intercepted by the in situ spacecraft. The large error in the speed, during collision phase of the CMEs, result in a severe limitation in the calculation of coefficient of restitution in the present study.

From Figure~\ref{insitu}, the region at the rear edge of CME2 during October 1, 18:00 UT to October 2, 02:00 UT shows enhancement in density up to 25 cm$^{-3}$ and depression in temperature up to 2 $\times$ 10$^{4}$. This region is identified as the cold filament, as the CME2 was associated with eruption of a filament. During the arrival of this cold and dense structure, the alpha to proton ratio increased up to 8\%, thermal speed of helium ion dropped to 20 km s$^{-1}$, average charge state of iron drops up to 9, and C$^{+6}$/C$^{+5}$ ratio decreased to 0.6. These findings further confirms the identification of filament materiel at 1 AU at the rear edge of CME2 as reported by earlier authors \citep{Burlaga1998,Gopalswamy1998,Lepri2010,Sharma2012}.

On the arrival of CME1, the southward component of magnetic field goes up to only -5 nT for 10 hr  leading to the development of  ring current in the Earth's magnetosphere. This is manifested as depression in the Earth's horizontal component of magnetic field known as geomagnetic storm and measured as depression in the Sym-H index shown at the bottom panel of the Figure~\ref{insitu}. Further, depression in Sym-H index takes place on the arrival of compressed region bounded between second and third vertical lines in the figure. Such a two step development of geomagnetic storm is expected because of  the arrival of the interacting CMEs. This finding is in agreement with \citet{Farrugia2006} that interacting CMEs having two ejecta can lead to intense two step geomagnetic storms as in the present case $Dst$ index reached up to -119 nT.

In the present case, the two CMEs although had started merging close to the Earth,  their signature of interaction/collision are different from studies reported earlier \citep{Mishra2015}. In this present case, the  interaction region, bounded between second and third vertical lines in Figure~\ref{insitu}, did not  reveal the  presence of a magnetic hole, shown by depressed magnetic field and enhanced temperature as signatures of reconnection. However, such interaction region was observed for the November 9-10 interacting CMEs \citep{Mishra2015}. Distinct structures corresponding to colliding CMEs have been shown for interacting CMEs of 2011 February \citep{Mishra2014a}. From these studies, it is expected that in the case of elastic collision, the interacting CMEs may be identified individually in the in situ observations at 1 AU. In the present study, various sources of uncertainties because of idealistic assumptions while estimating the kinematics, understanding the nature of collision and in situ measurements of interacting CMEs cannot be excluded. The discussion on such limitations is presented in detail in earlier studies \citep{Temmer2012,Mishra2014a,Mishra2015}.

\section{Summary}
We discuss the difficulty in reliably estimating the kinematics from SSE method for the Earth-directed CME when \textit{STEREO} is located more than 90$^\circ$ away the Sun-Earth line. For such a scenario, the importance of chosen value of $\lambda$ and propagation direction of CME in the SSE method is highlighted. We note the overestimation of speed of a far-sided CME, irrespective of its geometry used in SSE method. Our present study of interacting CMEs of  September 25 and 28, 2012 shows that the collision is close to elastic in nature and huge momentum exchange and kinetic energy change took place during the interaction. The interaction region formed close to 1 AU is unique as it is the region formed due to passage of shock associated with following CME through preceding CME. This region is responsible for enhanced geomagnetic activity. We also underline the difficulty in authentic calculation of coefficient of restitution when collision of CMEs takes place at larger distance from the Sun where estimated kinematic parameters have extreme uncertainties. The present study highlights the importance of wide-angle imaging enabling us to witness a unique  interaction close to the Earth.

\begin{acknowledgments}
We acknowledge the team members of \textit{STEREO} COR and HI as well as WIND instruments. We acknowledge the UK Solar System Data Center for providing the processed Level-2 \textit{STEREO}/HI data. The authors also thank Yuming Wang (USTC, China) and  Jackie A. Davies (RAL, UK) for useful discussion during the development of this work. We also thank the reviewers for their valuable comments and insight.  W.M. is supported by the Chinese Academy of Sciences President’s International Fellowship Initiative (PIFI) grant No. 2015PE015 and NSFC grant Nos. 41131065 and 41574165. 
\end{acknowledgments}

\appendix
\appendixpage
\section{Relationship between angular width of the CME as adopted for SSE and GCS model}
\label{lmbdwidth}
CME  angular extent approximated in SSE or SSSE model is measured by the $\lambda$ parameter which can be derived using measured 3D widths of the CME from GCS model.  Based on the fitted GCS model parameters (Section~\ref{3DCOR}), we first estimated the 3D face-on ($\omega_{FO}$) and  edge on ($\omega_{EO}$)  angular width of the CMEs \citep{Thernisien2011}.  For CME1, $\omega_{EO}$ = 2$\delta$ = 40$^{\circ}$, where $\delta$ = $\arcsin(\kappa$)  and for CME2,  $\omega_{EO}$ = 62$^{\circ}$ is estimated.  For CME1,  $\omega_{FO}$ = 2($\alpha$+$\delta$) = 82$^{\circ}$ and for CME2,  $\omega_{FO}$ is estimated as 199$^{\circ}$ .  The calculated 3D angular widths ($\omega$) are different from the value of $\lambda$ assumed in the SSSE method. This is because, $\omega$ is derived assuming the CME as a GCS structure while, $\lambda$ is for the CME assumed as a sphere (or circle in ecliptic plane).  Assuming GCS and SSE geometry in the ecliptic plane, our GCS fitting suggests that CME1 has a tilt angle ($\gamma$) of about 6$^\circ$, implying that its $\lambda$ value is more close to half of its face-on width. The tilt angle of CME2 is about -75$^\circ$ and this implies that its $\lambda$ value is close to the half of its edge-on width.  Hence, the $\lambda$ value of CME1 and CME2 is estimated as 41$^{\circ}$ and 31$^{\circ}$ respectively. The errors in GCS fitted parameters can result in error in the $\lambda$ value leading to erroneous estimation of CME structure assumed for SSE model. The error in $\lambda$ value of the CMEs is found to be around $\pm$10$^{\circ}$. 

\section{Effect of $\lambda$ on estimated distance from SSE method}
\label{lmbdSSE}
\begin{figure}[!htb]
\captionsetup{width=0.85\textwidth}
\begin{center}
\includegraphics[scale=0.6]{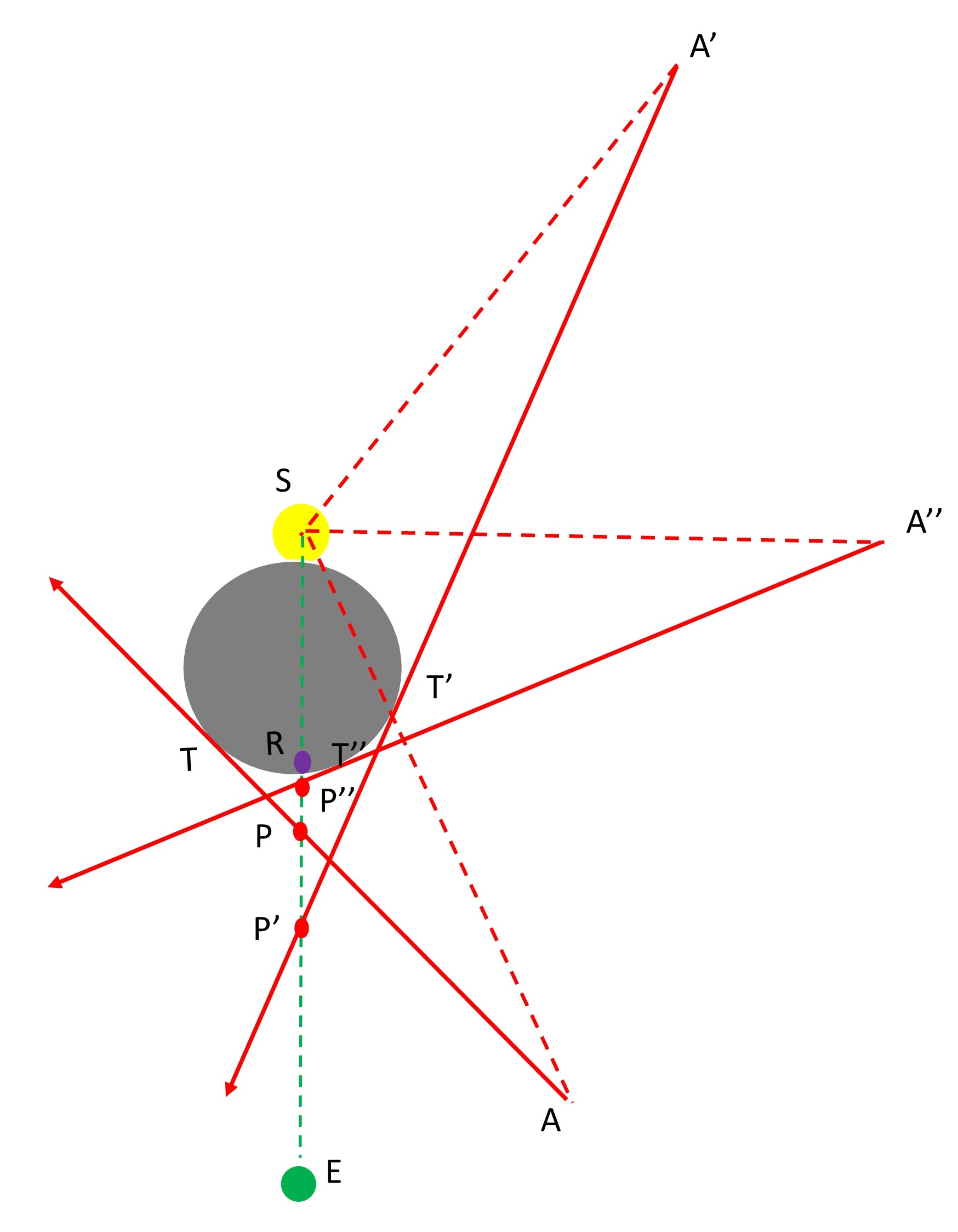}
\caption{In the figure, the Sun (S), Earth (E), and observers (A,A' and A'') are labeled.  A CME shown as a shaded circle directed towards the Earth is viewed by three observers located at different heliographic longitudes; near, perpendicular and away from the Sun-Earth line. T,T' and T'' are the points on CME where line-of-sight from  A, A' and A'' meet tangentially and R is the identified location of CME apex with HM method. P, P' and P'' are the locations where the CME apex will be identified from  A, A'  and A'' with Fixed-Phi method.}
\label{loca}
\end{center}
\end{figure}

\begin{figure}[!htb]
\captionsetup{width=0.85\textwidth}
\begin{center}
\includegraphics[angle=0,scale=0.80]{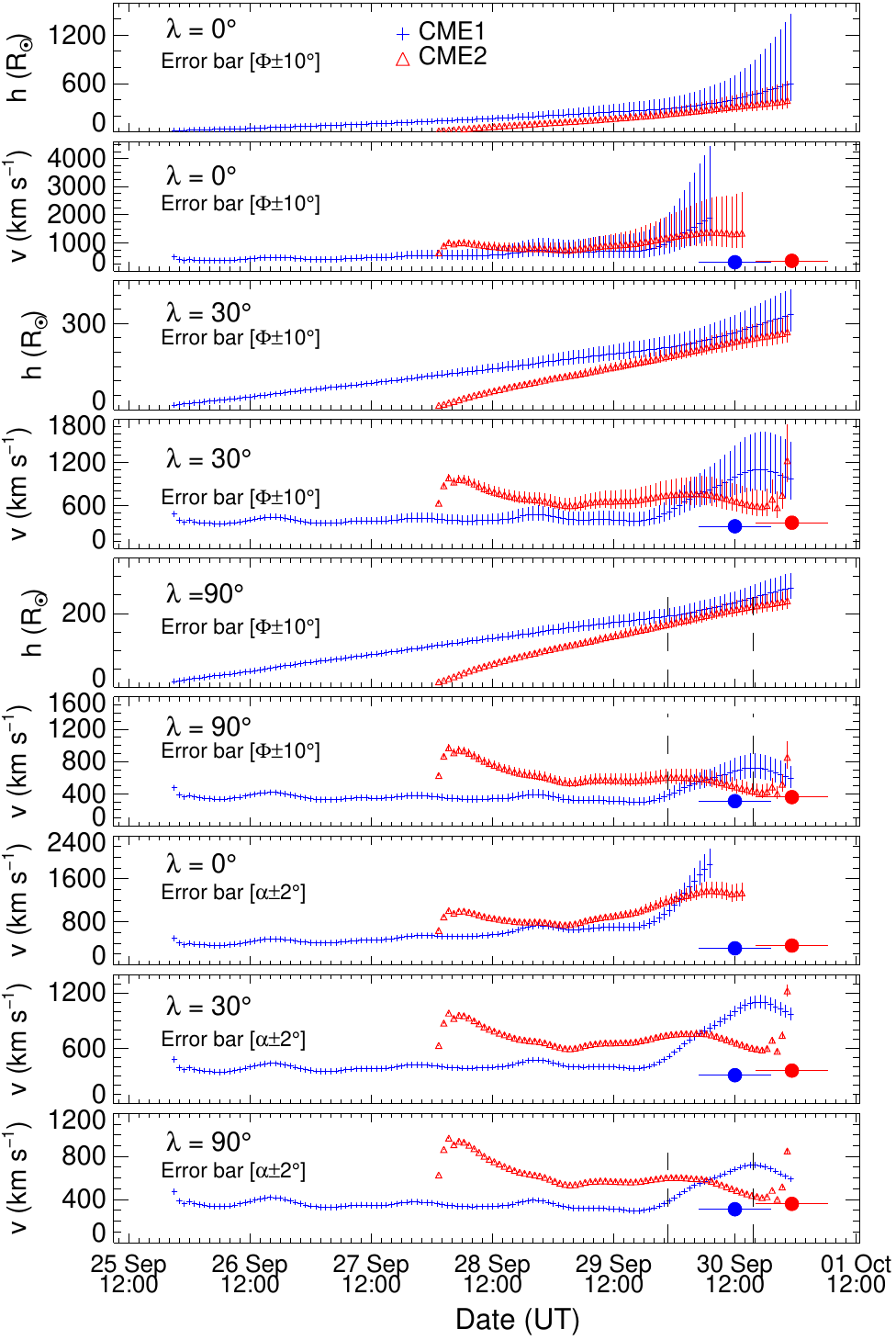}
\caption{3D kinematics of the interacting CMEs of September 25 and 28, 2012 using the SSE method. From the top; the first, third and fifth panels show evolution of 3D height while second, fourth and sixth panels show the evolution of 3D speed of the leading edge of the two CMEs corresponding to three different values of $\lambda$ = 0$^{\circ}$, 30$^{\circ}$, 90$^{\circ}$. In the top six panels, vertical lines show the errors bars, calculated using propagation directions ($\Phi$) that are +10$^\circ$ and -10$^\circ$ different to the value estimated using GCS model of 3D reconstruction. The error bars in the bottom three panels, for $\lambda$ = 0$^{\circ}$, 30$^{\circ}$, 90$^{\circ}$, represents the uncertainties in the kinematics because of using elongation angle ($\alpha$) that are $\pm$2$^{\circ}$ different than the observed for the CMEs. The horizontal lines and filled circles in the speed panels, respectively represent the in situ measured speed and arrival times of the CMEs. The vertical dashed lines, in panels for $\lambda$=90$^{\circ}$, marks the start and end of the collision phase.}
\label{HT_difflambda}
\end{center}
\end{figure}

\begin{figure}[!htb]
\captionsetup{width=0.85\textwidth}
\begin{center}
\includegraphics[angle=0,scale=0.85]{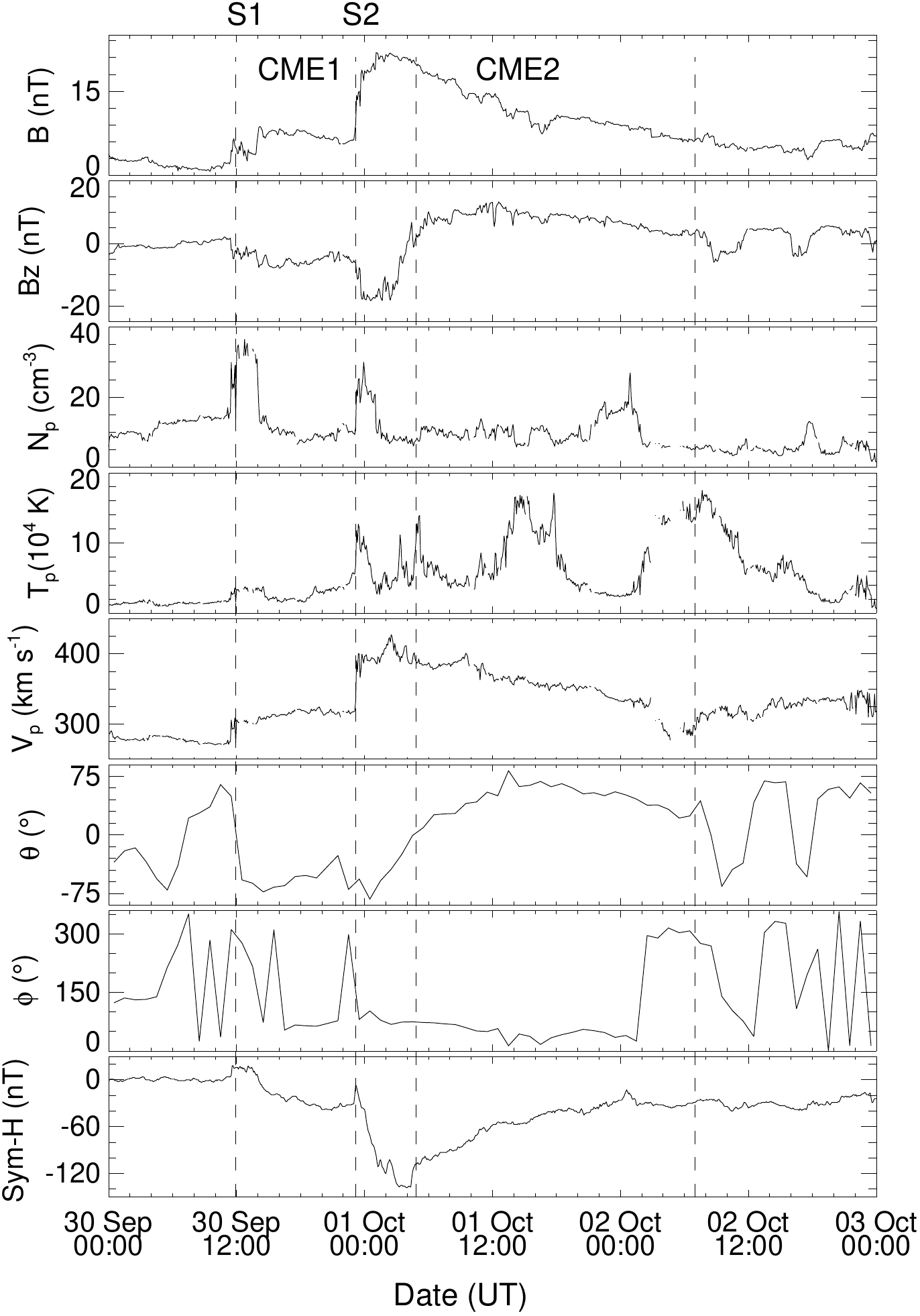}
\caption{In situ measurements of plasma and magnetic field parameters associated with the CMEs of September 25 and 28, 2012 during the period September 30 to October 3, 2012. S1 and S2 mark the arrival of shocks associated with the CME1 and CME2 respectively. The region of CME1 compressed by the S2 is shown between the second and third vertical dashed lines (from the left). The region at the rear edge of CME2 arrived during October 1, 18:00 UT to October 2, 02:00 UT corresponds to possible location of the eruptive filament associated with the CME2.}
\label{insitu}
\end{center}
\end{figure}

In Figure~\ref{loca}, S, E, A, A' and A'' represent the locations of the Sun, Earth, different position of \textit{STEREO-A} (observer) near, away and perpendicular to the Sun-Earth line, respectively. The line-of-sight  APT, A'T'P'  and A''T''P'' shown with solid lines observe the CME (gray shaded circle) at exactly a tangent. Under the HM approximation for a CME, an observer at A (or A' and A'') receives the signal from T (or T' and T'')  and then the distance (SR) of CME propagating towards the Earth can be estimated. However, under the FP approximation, the estimated height of CME will be SP, SP' and SP'' corresponding to observers position at A, A' and A''. Hence, for an observer at A in Figure~\ref{loca}, the difference in estimated heights (SP-SR) of the CME at intermediate elongation angle, from FP and HM methods is significantly less than its value (SP'-SR) for observer's position at A'. Mathematically, the measured distances of the CME from the Sun using FP  and HM methods can be written by equation~\ref{eqfp} and ~\ref{eqhm} \citep{Kahler2007,Lugaz2009}, respectively; where $d$ is the distance of observer from the Sun, $\alpha$ is the elongation (i.e. Sun-CME-observer angle) and $\Phi$ is the propagation angle of CME relative to Sun-observer line. These equations are special cases of  equation~\ref{eqsse} for SSE method \citep{Davies2012} corresponding to $\lambda$=0$^{\circ}$ and 90$^{\circ}$, respectively. These equations are derived for $\Phi$ greater than zero and less than 180$^{\circ}$. 
 
\begin{equation}
r_{FP}  = \frac{d\sin(\alpha)}{\sin(\alpha+\Phi)}
\label{eqfp}
\end{equation}

\begin{equation}
r_{HM}  = \frac{2d\sin(\alpha)}{1+\sin(\alpha+\Phi)}
\label{eqhm}
\end{equation}

\begin{equation}
r_{SSE}  = \frac{d\sin(\alpha)(1+\sin(\lambda))}{\sin(\alpha+\Phi)+\sin(\lambda)}
\label{eqsse}
\end{equation}

Using equation~\ref{eqfp} and ~\ref{eqhm}, we can write
\begin{equation}
\frac{r_{FP}}{r_{HM}} = \frac{1+\mathrm{cosec}(\alpha+\Phi)}{2}
\label{rfphm}
\end{equation}

The value of a trigonometric cosec function of an angle ($\Theta$), for 0$^{\circ}$ $<$ $\Theta$ $<$180$^{\circ}$, is always positive and greater than 1 (except at $\Theta$=90$^{\circ}$), implying that the distance measured from FP method ($r_{FP}$) is greater than the distance measured from HM method ($r_{HM}$). The value of cosec($\Theta$) decreases for 0$^{\circ}$ to 90$^{\circ}$ and this decrease is very fast up to $\approx$ 30$^{\circ}$) and then slowly reaches to 1 at 90$^{\circ}$. However,  cosec($\Theta$) value increases for 90$^{\circ}$ to 180$^{\circ}$ and the increase is slow up to $\approx$ 150$^{\circ}$ and beyond this is extremely fast. From this trend of variation, it is noted that when  $\Phi$ $<$ 90$^{\circ}$ (e.g. observer at A in Figure~\ref{loca}) in equation~\ref{rfphm}, the difference between distance measured from FP and HM method first decreases and then increases with increasing elongation angle ($\alpha$). The difference becomes zero when  $\alpha$+$\Phi$ becomes 90$^\circ$ and further increases slowly with increasing elongation angle. However, for $\Phi$ $>$90$^{\circ}$ (e.g. observer at A' in Figure~\ref{loca}), the difference between distance measured from FP and HM method always increases with increasing elongation angle. Hence, for an observer location approximately perpendicular to CME propagation direction (i.e. at A'' in Figure~\ref{loca}), both HM and FP methods can measure almost the same distance for smaller elongation angle. We emphasize that a large discrepancy between the results from FP and HM methods is noted for $\Phi$$<$30$^{\circ}$ and $\Phi$$>$150$^{\circ}$. Such discrepancy is progressively reduced with increasing elongation angle if the CME is not far-sided for an observer.

\section{Overestimation of CME speed}
\label{lmbdacc}
We can infer from Equation~\ref{eqfp} that when the CME propagates largely away from the observer (i.e. 90$^{\circ}$$<$$\Phi$$<$180$^{\circ}$), the value of  $\sin(\alpha+\Phi)$ in the denominator decreases much faster than increase of $\sin(\alpha)$ in the numerator, on increasing the value of $\alpha$. This leads to an unphysical large acceleration for a far-sided CME at higher elongation irrespective of the value of $\lambda$. However, such acceleration is significantly reduced using Equation~\ref{eqhm}. This is because, in Equation~\ref{eqhm} the numerator is twice and denominator is unity larger than the numerator and denominator of Equation~\ref{eqfp}. Such finding is also shown by \citet{Liu2013}.  For $\lambda$=30$^{\circ}$ in Equation~\ref{eqsse}, its numerator becomes 1.5 times and denominator becomes a half larger than the numerator and denominator of Equation~\ref{eqfp}. Hence, such unphysical acceleration will also be reduced using Equation~\ref{eqsse} for SSE method, however not up to the extent as from HM method. It is noted that all the single spacecraft methods, irrespective of the value of $\lambda$ used in the SSE method, overestimate the speed of  far-sided CMEs. The flattening of CME front due to its interaction with solar wind also partially contributes in overestimation of CME speed from SSE method.   

%

\end{article}

\clearpage

\end{document}